\documentclass[sn-mathphys]{sn-jnl}

\jyear{2022}%

\theoremstyle{thmstyleone}%
\newtheorem{theorem}{Theorem}
\newtheorem{proposition}[theorem]{Proposition}%

\theoremstyle{thmstyletwo}%
\newtheorem{example}{Example}%
\newtheorem{remark}{Remark}%

\theoremstyle{thmstylethree}%

\usepackage{graphicx}
\usepackage{amsmath,amssymb,amsfonts,amsthm}
\usepackage[version=4]{mhchem}
\usepackage{siunitx}
\usepackage{hyperref}
\usepackage{cleveref}
\usepackage{longtable,tabularx}
\usepackage{mathtools}
\usepackage{mdwlist}
\usepackage{verbatim}
\usepackage{textcomp}
\let\savelneq\lneq
\let\lneq\relax
\usepackage{mathabx}
\let\lneq\savelneq

\usepackage{graphicx}
\usepackage{etoolbox}
\usepackage{enumitem}

\newcommand{\bu}{\textbf{\emph{u}}}
\newcommand{\bv}{\textbf{\emph{v}}}

\newcommand{\bs}{\textbf{\emph{s}}}
\newcommand{\bg}{\textbf{\emph{g}}}
\newcommand{\bc}{\textbf{\emph{c}}}
\newcommand{\brr}{\textbf{\emph{r}}}
\newcommand{\bp}{\textbf{\emph{p}}}
\newcommand{\bq}{\textbf{\emph{q}}}
\newcommand{\bw}{\textbf{\emph{w}}}
\newcommand{\bx}{\textbf{\emph{x}}}

\newcommand{\bC}{\textbf{\emph{C}}}

\newcommand{\bT}{\textbf{\emph{T}}}

\newcommand{\bP}{\textbf{\emph{P}}}
\newcommand{\bH}{\textbf{\emph{H}}}
\newcommand{\bR}{\textbf{\emph{R}}}
\newcommand{\bQ}{\textbf{\emph{Q}}}

\newcommand{\bS}{\textbf{\emph{S}}}

\newcommand{\bU}{\textbf{\emph{U}}}
\newcommand{\bV}{\textbf{\emph{V}}}
\newcommand{\bI}{\textbf{\emph{I}}}
\newcommand{\bA}{\textbf{\emph{A}}}
\newcommand{\bF}{\textbf{\emph{F}}}
\newcommand{\bX}{\textbf{\emph{X}}}

\newcommand{\bl}{\boldsymbol{\ell}}
\newcommand{\bpi}{\boldsymbol{\pi}}
\newcommand{\Qerr}{\Delta \bQ^*}
\newcommand{\aerr}{\Delta a}
\newcommand{\eerr}{\Delta e}
\newcommand{\ierr}{\Delta i}
\newcommand{\longerr}{\Delta \Omega}
\newcommand{\argerr}{\Delta \omega}
\newcommand{\earthrad}{\mathbf{R}_{\oplus}}

\newcommand{\PP}{{\mathbb{P}}}
\newcommand{\RR}{{\mathbb{R}}}
\newcommand{\CC}{{\mathbb{C}}}

\newcommand{\rvline}{\hspace*{-\arraycolsep}\vline\hspace*{-\arraycolsep}}
\DeclareMathOperator*{\argmin}{arg\,min}
\DeclareSIUnit\earthradius{\earthrad}
\theoremstyle{definition}
\AtBeginEnvironment{proposition}{\setlist{before=\leavevmode, nosep}}
\newcommand{\hide}[1]{}

\begin{document}

\title[Geometric Solution to the Angles-Only Initial Orbit Determination Problem]{Geometric Solution to the Angles-Only Initial Orbit Determination Problem\footnote{An earlier version of this manuscript was presented as Paper AAS 22-687 at the AAS/AIAA Astrodynamics Specialist Conference in August 2022.}}

\author*[1]{\fnm{Michela} \sur{Mancini}}\email{mmancini32@gatech.edu}

\author[2]{\fnm{Timothy} \sur{Duff}}

\author[3]{\fnm{Anton} \sur{Leykin}}

\author*[1]{\fnm{John A.} \sur{Christian}}\email{john.a.christian@gatech.edu}

\affil[1]{\orgdiv{Guggenheim
School of Aerospace Engineering}, \orgname{Georgia Institute of Technology}, \city{Atlanta}, \postcode{30332}, \state{GA}, \country{USA}}

\affil[2]{\orgdiv{Department of Mathematics}, \orgname{University of Washington}, \city{Seattle}, \postcode{98195}, \state{WA}, \country{USA}}

\affil[3]{\orgdiv{School of Mathematics}, \orgname{Georgia Institute of Technology}, \city{Atlanta}, \postcode{30332}, \state{GA}, \country{USA}}

\abstract{Initial orbit determination (IOD) from line-of-sight (i.e., bearing) measurements is a classical problem in astrodynamics. Indeed, there are many well-established methods for performing the IOD task when given three line-of-sight observations at known times. Interestingly, and in contrast to these existing methods, concepts from algebraic geometry may be used to produce a purely geometric solution. This idea is based on the fact that bearings from observers in general position may be used to directly recover the shape and orientation of a three-dimensional conic (e.g., a Keplerian orbit) without any need for knowledge of time. In general, it is shown that five bearings at unknown times are sufficient to recover the orbit---without the use of any type of initial guess and without the need to propagate the orbit. Three bearings are sufficient for purely geometric IOD if the orbit is known to be (approximately) circular. The method has been tested over different scenarios, including one where extra observations make the system of equations over-determined. 
}




\maketitle
\section{Introduction}
Initial orbit determination (IOD) is one of the classical problems in astrodynamics. The problem has attracted the attention of many notable mathematicians---including Gauss \cite{Gauss:1809}, Laplace \cite{Laplace:1780}, and Gibbs \cite{Gibbs:1889}---whose algorithms are still in widespread use hundreds of years later. The various classical IOD algorithms address different IOD scenarios, with each scenario having its own set of unique assumptions about the information available for the orbit determination. Some examples are highlighted in~\Cref{tab:classicODmethods}.

Amongst the classical problems (first three rows) from Table~\ref{tab:classicODmethods}, the Gibbs problem is unique in that it is purely geometric. Specifically, it finds the Keplerian orbit (a 3D conic with focus at the origin) that passes through three known points in space---without any need for explicitly knowing the time at which the orbiting object resided at those three points. Similar geometric solutions have also been identified for the velocity-only IOD problem \cite{Hollenberg:2019} through use of the orbital hodograph (fourth row in Table~\ref{tab:classicODmethods}). To our knowledge, no such purely geometric solution exists for IOD from only line-of-sight (i.e., bearing) measurements---the so-called ``angles-only IOD'' scenario. Popular angles-only IOD algorithms---including those of Gauss \cite{Gauss:1809}, Laplace \cite{Laplace:1780}, Gooding \cite{Gooding:1997}, and others \cite{Escobal:1976}---all require the use of time to propagate the body between specific points on the orbit (corresponding to the bearing measurements). Therefore, in this work we apply concepts from algebraic geometry to produce a purely geometric solution to the angles-only IOD problem. More specifically, we formulate the IOD problem in terms of solving a system of multivariate polynomials.
Solving this system of polynomials for a minimum of five observations (or three in the case of a circular orbit) reduces estimating the unknown orbital parameters to checking finitely many candidates.

\begin{table}[!ht]
\caption{Comparison of some common IOD problem formulations.}\label{tab:classicODmethods}
\centering
\begin{tabular}{l|c|c|c}
\hline
\hline
 & \textbf{Observation} & \textbf{Number of} & \textbf{Explicit Use} \\
 & \textbf{Type} & \textbf{Observations} & \textbf{of Time}\\
 \hline
 Angles-Only (Gauss, Laplace &  \multirow{2}{*}{Bearings} & \multirow{2}{*}{3} & \multirow{2}{*}{Y}\\
 Double-R, \& Gooding) \cite{Escobal:1976} & & & \\
 \hline
 Lambert \cite{Lambert} & Position & 2 & Y\\
 \hline
 Gibbs \cite{Gibbs:1889,Escobal:1976} & Position & 3 & N\\
 \hline 
 Hodograph (Vectors) \cite{Hollenberg:2019} & Velocity & 3 & N\\
 \hline
 Hodograph (Angles) \cite{Christian:2023}  & Headings & 4 & Y\\
 \hline
 This Work & Bearings & 3 or 5 & N\\ 
\hline
\end{tabular}
\end{table}

The remainder of this paper develops in the following way. We first review some key concepts from algebraic geometry (e.g., dual representation of quadrics, lines, and planes) and relate these ideas to the problem of two-body orbital mechanics and bearing observations. These ideas are used to recast the IOD problem as a system of polynomials. We introduce a solution method based on the idea of homotopy continuation. The solution is found to be straightforward within this framework and we demonstrate the efficacy on a number of example orbits, including a nearly circular orbit, a highly elliptical orbit, and a hyperbolic orbit.

\section{Geometry of Keplerian Orbits and Line-of-Sight Measurements}

The angles-only IOD may be solved without using the observation times by approaching the problem from a geometric (rather than dynamical) standpoint. Such a geometric interpretation leads us think of the orbit as a space conic, which may be compactly represented in terms of its disk quadric. Thus, our development begins by considering we might relate classical orbital elements to the disk quadric and what orbit parameterizations are most desirable. 

Consider a particle (e.g., spacecraft, celestial body) in an unknown orbit that we wish to estimate. Under the assumption of Keplerian motion, the spacecraft will move along a path that is a conic section. The plane of this conic section is constrained to pass through the center of the gravitating body (taken to be the origin), though the plane's orientation is unknown. Moreover, one of the conic section's foci must lie at the origin.

\begin{figure}[ht!]
\centering
\includegraphics[width=0.5\columnwidth,trim=0in 0in 0in 0in,clip]{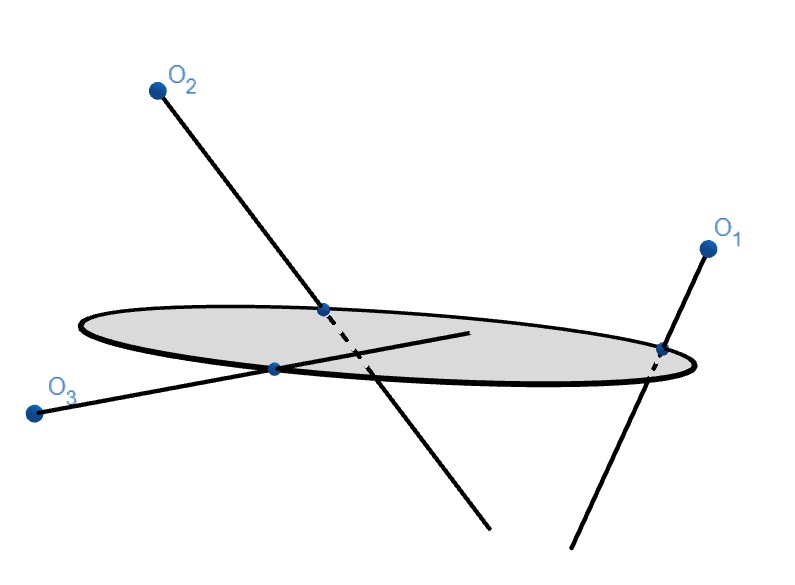}
	\caption{Illustration of geometry. The points \(O_i\) represent the observers' positions and the lines the observed lines.}
	\label{fig:ProblemGeomOverview}
\end{figure}

Now, suppose we have a set of $n$ observations of this orbiting particle from $n$ different observers in general position. Each observation consists of the bearing (i.e., the line-of-sight direction) from the observer's location to the particle's location. Thus, we have a set of $n$ lines, with each line passing through the conic formed by the spacecraft orbit and one of the observer points. This is illustrated in Fig.~\ref{fig:ProblemGeomOverview}.

The physical constraint that we want to impose is that the intersection between the lines of sight and the orbit happens. By reformulating the IOD problem in this way we are able to free the solution from inaccuracies related to corrections for the light time-of-flight that are usually needed when associating a measurement to a time instant, and that may lead to errors whose extent is often unpredictable \textit{a priori}.

\subsection{Keplerian Orbit as a Disk Quadric}
A Keplerian orbit is a 3D space conic, which may be described in many ways.
For instance, it may be represented as the intersection of a plane with a quadric surface, e.g.~a cone or an ellipsoid. 
A less cumbersome description is the \emph{dual representation} of the conic, which describes its locus of tangent planes.
In two dimensions this is sometimes called the conic envelope, and in 3D it is usually called the \emph{disk quadric} \cite{Semple:1952}. The reader interested in additional details on quadrics is directed to Refs.~\cite{Semple:1952,Hartley:2003,Christian:2021}.

Since it is both convenient and natural to work with homogeneous coordinates, we will regard a plane in space as a subset of the three-dimensional projective space $\PP^3.$
A plane in $\PP^3$ may be represented either in \emph{primal} terms, in which case the points on the plane are vectors in the column span of a $4\times 3$ matrix, or in \emph{dual} terms as the left-nullspace of $4\times 1$ matrix $\bpi .$ 
Similarly, a line in $\PP^3$ may be represented in \emph{primal} terms as the column span of a $4\times 2$ matrix, or in \emph{dual} terms, in which case the planes containing the line are vectors in the column span of a $4\times 2$ matrix.  

The disk quadric may be represented by a rank-deficient $4\times 4$ matrix $\bQ^*$.
In dual terms, a plane $\bpi $ lies on the disk quadric if and only if
\begin{equation}\label{eq:disk-quadric}
\bpi^T \bQ^* \bpi = 0.
\end{equation}
which describes a surface in dual projective space.
Intuitively, the disk quadric may be viewed as all the planes tangent to an ellipsoid that has been flattened into the shape of a pancake.

\subsection{Orbit as a Disk Quadric}
\begin{figure}[ht!]
\centering
\includegraphics[width=0.5\columnwidth,trim=0in 0in 0in 0in,clip]{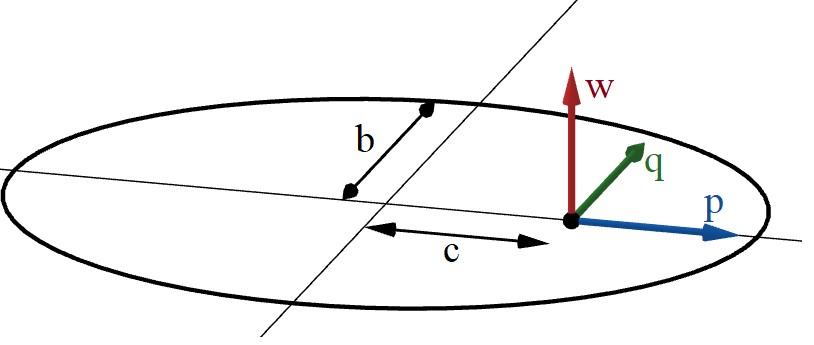}
	\caption{The semi-minor axis \(b\) and the focal length \(c\) completely characterize the orbit in the orbital plane. The \(\bp\) and \(\bq\) axes determine the orientation of the orbit in space.}
	\label{fig:PerifocalFrame}
\end{figure}

We will proceed by relating the disk quadric $\bQ^*$ to more commonly used orbital elements. Therefore, consider an orbit whose perifocal frame is given by the orthonormal basis vectors $\{\bp,\bq,\bw\}$. We choose the convention where $\bp$ points from the origin (located at the center of the gravitating body) to the orbit periapsis and where $\bw$ is normal to the orbit plane (in the direction of the angular momentum vector). The unit vector $\bq$ completes the right-handed system. This frame is shown in Fig.~\ref{fig:PerifocalFrame}.

If the orbit has a semi-major axis of $a$ and eccentricity of $e$, then the distance from the origin (which must lie at one of the ellipse foci) to the orbit geometric center is given by $c = e a$. Likewise, we recall that the semi-minor axis $b$ is related to $a$ and $c$ according to $a^2 = b^2 + c^2$. Thus, in the perifocal frame, one may write the ellipse as
\begin{equation}
    \frac{(x+c)^2}{a^2} + \frac{y^2}{b^2} = 1,
\end{equation}
or, equivalently,
\begin{equation}
b^2 x^2 + 2 b^2 c x + b^2 c^2  + a^2 y^2 - a^2 b^2 = 0.
\end{equation}
Collecting terms, we may write
\begin{equation}
b^2 x^2 + a^2 y^2 + 2 b^2 c x + b^2 (c^2-a^2) = 0,
\end{equation}
\begin{equation}
b^2 x^2 + a^2 y^2 + 2 b^2 c x - b^4 = 0.
\end{equation}
In homogeneous coordinates $\bx^T \propto \, [x,y,1]$, this last equation is equivalent to
\begin{equation}
    \bx^T \bC \bx = 0,
\end{equation}
where $\bC$ is a nonzero $3\times 3$ matrix given up to scale:
\begin{equation}
    \bC \, \propto 
    \begin{bmatrix}
        b^2 & 0 & b^2 c \\
        0 & a^2 & 0 \\
        b^2 c & 0 & -b^4
    \end{bmatrix}.
\end{equation}
If $\bC$ describes the conic locus of the orbit, then $\bC^* \propto \, \bC^{-1}$ describes the conic envelope (lines tangent to the conic in the orbit plane). Therefore, recalling $a^2 = b^2 + c^2$,
\begin{equation}
    \bC^* \propto \, \bC^{-1} = 
    \begin{bmatrix}
        1/a^2 & 0 & c/(a^2 b^2) \\
        0 & 1/a^2 & 0 \\
        c/(a^2 b^2) & 0 & -1/(a^2 b^2)
    \end{bmatrix} \propto
    \begin{bmatrix}
        1 & 0 & c/b^2 \\
        0 & 1 & 0 \\
        c/b^2 & 0 & -1/b^2
    \end{bmatrix}.
\end{equation}

The objective now is to relate the conic envelope to the disk quadric (a type of quadric envelope). To do this, we recall from Ref.~\cite{Christian:2021crater} the relation
\begin{equation}
    \bQ^* \propto \, \bH \bC^* \bH^T
\end{equation}
where, in this case, one may compute $\bH$ as the $4 \times 3$ matrix
\begin{equation}
    \bH = 
    \begin{bmatrix}
       \bp & \bq  & \textbf{0}_{3 \times 1} \\
       0 & 0  & 1
    \end{bmatrix}.
\end{equation}
Performing the requisite multiplications gives a disk quadric of 
\begin{align}
    \bQ^* \propto \, \bH \bC^* \bH^T 
        = \begin{bmatrix}
                \bp \bp^T + \bq \bq^T  & (c/b^2) \bp \\
                (c/b^2) \bp^T  & (-1/b^2)
            \end{bmatrix}.
\end{align}
Thus,
\begin{equation}
\label{eq:qstar}
\bQ^* \propto 
\begin{bmatrix}
\bp \bp^T + \bq \bq^T & (c/b^2) \bp \\
(c/b^2) \bp^T & (-1/b^2)
\end{bmatrix} = 
\begin{bmatrix}
\bI_{3 \times 3} - \bw \bw^T & (c/b^2) \bp \\
(c/b^2) \bp^T & (-1/b^2)
\end{bmatrix}.
\end{equation}
Consequently, the $4 \times 4$ matrix $\bQ^*$ describing the disk quadric consists of two scalars ($b$ and $c$, which describe the size and shape of the orbit in the orbital plane) and the two orthonormal vectors that span the perifocal plane ($\bp$ and $\bq$). Since $\bp$ and $\bq$ are orthonormal, we have the three constraints
\begin{equation}
\label{eq:constraints-pq}
\bp^T \bp = \bq^T \bq = 1, \phantom{ff}
\bp^T \bq = 0.
\end{equation}

The two parameterizations of $\bQ^{\ast}$ from Eq.~\eqref{eq:qstar} and the three constraints from Eq.~\eqref{eq:constraints-pq} were also given without derivation in Ref.~\cite{Christian:2021streak}. The difficulty with these specific ways of writing $\bQ^{\ast}$ is that they require different treatment for an elliptical orbit (when the direction of $\bp$ is well-defined) and for a circular orbit (when the direction of $\bp$ is \emph{not} well-defined). Thus, we introduce a third parameterization of $\bQ^{\ast}$ that avoids this deficiency. Specifically, define the vector $\bg = (c/b^2)\bp$, which is well-defined for both elliptical and circular orbits. We observe, in the limit as $e \rightarrow 0$, that $c \rightarrow 0$ and $\bg \rightarrow \textbf{0}_{3 \times 1}$. 
Therefore, we may parameterize the disk quadric as
\begin{equation}
\label{eq:qstar2}
\bQ^* \propto 
\begin{bmatrix}
\bI_{3 \times 3} - \bw \bw^T & \bg \\
\bg^T & (-1/b^2)
\end{bmatrix},
\end{equation}
where the parameters $\bw, \bg, b$
satisfy the two constraints
\begin{equation}
\label{eq:constraints-wg}
\bw^T \bw = 1, \phantom{ff}
\bw^T \bg = 0.
\end{equation}
Thus, because of its benefits for near-circular orbits, the parameterization of $\bQ^{\ast}$ from Eq.~\eqref{eq:qstar2} is used in the developments that follow.

\subsection{Line-of-Sight Observations and Tangent Planes}

\begin{figure}[ht!]
\centering
\includegraphics[width=0.5\columnwidth,trim=0in 0in 0in 0in,clip]{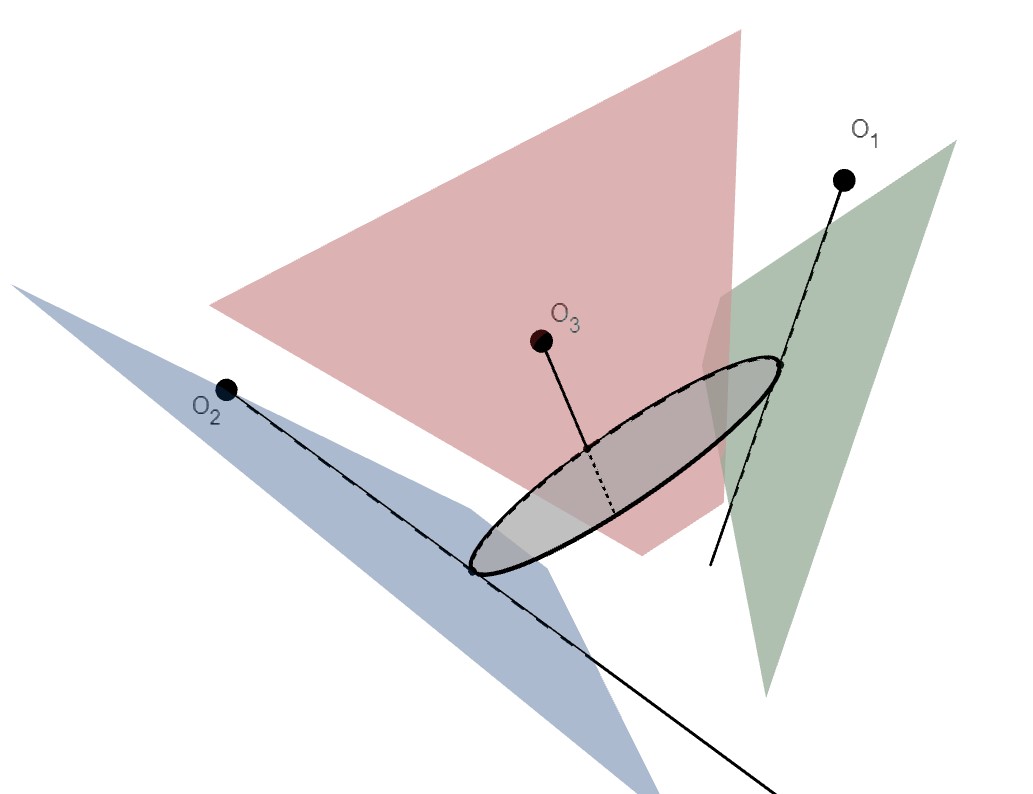}
	\caption{Representation of the planes $\bpi_1, \bpi_2, \bpi_3$ of the disk quadric containing three observed lines from the three observers \(O_1\), \(O_2\) and \(O_3\).}
	\label{fig:TangentPlanes}
\end{figure}

A plane in $\PP^3$, represented by $\bpi \in \RR^{4\times 1},$ is tangent to $\bQ$ if and only if
\begin{equation}
\label{eq:tangency}
\bpi^T \bQ^* \bpi = 0.
\end{equation}
Observed lines $\bl_1, \dots , \bl_n \subset \PP^3$ must lie within the unknown tangent planes $\bpi_1, \dots , \bpi_n,$ obtained geometrically as the join of each observed line with the corresponding tangent line in the orbit plane. Each $\bpi_i$ constructed in this way is the unique plane in $\bQ^{\ast}$ that contains $\bl_i$, as illustrated in Fig.~\ref{fig:TangentPlanes}.

Each observed line $\bl_i$ may be described in primal coordinates as the column span of the $4\times 2 $ matrix 
\begin{equation}
\begin{bmatrix}
\bx_i & \bu_i\\
1 & 0
\end{bmatrix}.
\end{equation}
Here $\bx_i \in \RR^3$ gives the coordinates of the observer point $O_i$ and $\bu_i \in \mathcal{S}^2$ a point on the unit sphere, representing a line-of-sight measurement.
The bearings, in noiseless scenarios, are unit vectors pointing from observers towards points on the orbit.

Dually, $\bl_i$ may be represented by \emph{any} full-rank $4\times 2$ matrix $\bA_i$ such that
\begin{equation}
\label{eq:DualLineWithA}
\bA_i^T \, \begin{bmatrix}
\bx_i & \bu_i\\
1 & 0
\end{bmatrix} =
\mathbf{0}_{2\times 2}.
\end{equation}

Many choices are possible for the matrix $\bA_i$ giving the dual representation of $\bl_i$.
In the next subsection, we explain two methods---one algebraic, and the other numerical---by which a suitable $\bA_i$ may be directly computed from observations.
Once $\bA_i$ is computed, the tangent plane $\bpi_i$ may be written as 
\begin{equation}
\bpi_i = \bA_i \bc_i
\end{equation}
for a $2\times 1$ vector $\bc_i.$
Substituting this expression into Eq.~\eqref{eq:tangency} yields
\begin{equation}
\bc_i^T \bA_i^T \bQ^* \bA_i \bc_i = 0.
\end{equation}
The nonzero vector $\bc_i$, which depends on the unknown tangent plane, may be eliminated from this equation to obtain a constraint depending only on the observed line and the unknown entries of the disk quadric. Specifically,
\begin{equation}\label{eq:det-constraint}
\det \left( \bA_i^T \bQ^* \bA_i \right) = 0.
\end{equation}   

Thus, each of the observed lines $\bl_1, \ldots , \bl_n$ places a constraint on the 3D conic that is polynomial in the entries of $\bA_i$ and $\bQ^*.$
These are the basic constraints from which we may determine the unknown orbit.

\subsubsection{Dual representation of a line-of-sight observation}
The dual representation of the line $\bl_i$ is given by the matrix $\bA_i$. To express $\bA_i$ algebraically in terms of the observer vector $\bx_i$ and the bearing vector $\bu_i,$
it is convenient to partition the primal matrix into blocks, $\bS_1, \bS_2 \in \RR^{2\times 2}$,
\begin{equation}
\begin{bmatrix}
\bx_i & \bu_i\\
1 & 0
\end{bmatrix} =
\begin{bmatrix}
\bS_1 \\
\bS_2 
\end{bmatrix}.
\end{equation}
Assuming the observations $\bx_i, \bu_i$ are sufficiently generic, then both $\bS_1$ and $\bS_2$ will be invertible.
Thus, application of Eq.~\eqref{eq:DualLineWithA} allows us to parameterize $\bA_i$ as 
\begin{equation}
\bA_i^T = \begin{bmatrix} 
\bA_i ' & 
\rvline 
& 
- \bA_i ' \, \bS_1 \, \bS_2^{-1}
\end{bmatrix},
\end{equation}
where $\bA_i'$ may be chosen to be any invertible $2\times 2$ matrix.
Write $\bx_i = \begin{bmatrix}x_{i,1} & x_{i,2} & x_{i,3} \end{bmatrix}^T$ and $\bu_i = \begin{bmatrix}u_{i,1} & u_{i,2} & u_{i,3} \end{bmatrix}^T$.  
Noting that $\det \bS_2 = u_{i,3}$, we find it convenient to make the choice
$\bA_i ' = u_{i,3} \, \bI_{2\times 2}.$ We then compute
\begin{equation}
\bA_i ' \, \bS_1 \,\bS_2^{-1} =
\begin{bmatrix} 
x_{i,1} & u_{i,1} \\
x_{i,2} & u_{i,2}
\end{bmatrix}
\begin{bmatrix} 
0 & - u_{i,3} \\
-1 & x_{i,3}
\end{bmatrix}
= \begin{bmatrix} 
- u_{i,1} & x_{i,3} u_{i,1} - x_{i,1} u_{i,3}\\
- u_{i,2} & x_{i,3} u_{i,2} - x_{i,2} u_{i,3}
\end{bmatrix},
\end{equation}
and hence
\begin{equation}\label{eq:algebraic-Ai}
\bA_i = 
\begin{bmatrix} 
u_{i,3} & 0 \\
0 & u_{i,3} \\
u_{i,1} & u_{i,2} \\
x_{i,1} u_{i,3} - x_{i,3} u_{i,1} &
x_{i,2} u_{i,3} - x_{i,3} u_{i,2}
\end{bmatrix}.
\end{equation}
A numerical alternative to the dual representation of Eq.~\eqref{eq:algebraic-Ai} may be obtained from the singular value decomposition (SVD) of the transposed primal matrix,
\begin{equation}
\begin{bmatrix}
\bx_i^T & 1\\
\bu_i^T & 0
\end{bmatrix}
= \bU_i \boldsymbol{\Sigma} \bV_i^T =
\bU_i \begin{bmatrix}
\sigma_{i,1} & 0 \\
0 & \sigma_{i,2} \\
0 & 0 \\
0 & 0
\end{bmatrix}
\begin{bmatrix}
\bv_{i,1} & \bv_{i,2} & \bv_{i,3} & \bv_{i,4}
\end{bmatrix}^T.
\end{equation}
Since $\bA_i$ must lie in the null space of the transposed primal matrix, we may select $\bA_i$ using the last two columns of $\bV_i$,
\begin{equation}\label{eq:numerical-Ai}
\bA_i = \begin{bmatrix}
\bv_{i,3} & \bv_{i,4}
\end{bmatrix}^T.
\end{equation}

\subsubsection{Geometric interpretation of the algebraic constraint}
The algebraic constraint from Eq.~\ref{eq:det-constraint} may be found by considering the projective geometry for a camera. Suppose the LOS measurements $\bu_i$ were obtained by cameras located at positions $\bx_i$. For each of these cameras, let \(\bT_i\) be the proper orthogonal matrix describing the rotation from the inertial frame to the camera frame.
The orbiting particle at the position \(\brr_i\) in the inertial frame will be imaged at the image plane coordinates \(\overline{\bs}_i\in\mathbb{P}^2\) \cite{Hartley:2003,Henry:2023}
\begin{equation}
    \overline{\bs}_i\,\propto\, \bT_i\left(\brr_i-\bx_i\right),
\end{equation}
If we let the projection matrix \(\bP_i\) be
\begin{equation}
\bP_i=\begin{bmatrix}
\bT_i &  \vline & -\bT_i \bx_i,
\end{bmatrix}
\end{equation}
we can also write the projection transformation as
\begin{equation}
    \overline{\bs}_i \, \propto \,\bP_i \begin{bmatrix}
    \brr_i\\
    1\\
    \end{bmatrix}.
\end{equation}

\begin{figure}[!ht]
\centering
\includegraphics[width=0.7\columnwidth,trim=0in 0in 0in 0in,clip]{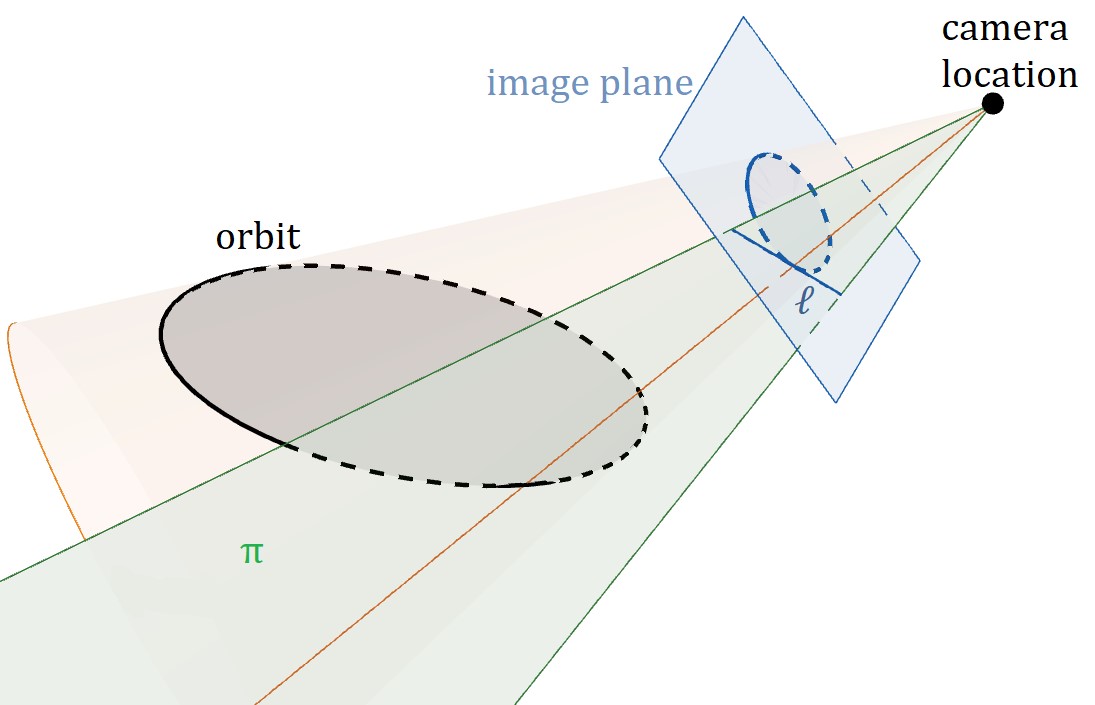}
	\caption{The plane of the disk quadric \(\bpi\) projects to the image plane in a line \(\boldsymbol{\ell}\), that is tangent to the projected conic.}
	\label{fig:DiskQuadricProjection}
\end{figure}

Now, imagine that the orbit as an actual conic in space. If we assemble all of the LOS directions originating from the camera and passing through the conic, we will obtain a cone having its vertex at the camera's location. A slice of this cone with the image plane creates another conic corresponding with the instantaneous projection of the entire orbital path into the image. Now, consider a plane tangent to the cone. By construction, this plane is also tangent to the orbit and, therefore, must be a plane belonging to the orbit's disk quadric. As shown in Fig.~\ref{fig:DiskQuadricProjection}, this plane intersects the image plane in a line that is tangent to the projected conic (i.e. is a line of the image plane conic envelope). Thus, we find that the disk quadric projects to a conic envelope \cite{Hartley:2003} that is a function of $\bQ^*$ (i.e., of the unknowns \(\bg, \bw\) and \(b\) ),
\begin{equation}
    \bC_i^* \, \propto \, \bP_i\bQ^*\bP_i^T.
\end{equation}

For a camera, each LOS measurement corresponds to an image plane measurement \(\overline{\bs}_i\) that must lie on the projected orbit's conic locus \(\bC_i \propto \, (\bC_i^*)^{-1}\), and thus satisfies the constraint:
\begin{equation}\label{eq:inplaneconstraint}
    \overline{\bs}^T_i \bC_i \overline{\bs}_i = 0.
\end{equation}
which is the conic locus equation. It is possible to verify that this polynomial constraint, once expanded and normalized, is equivalent to the algebraic constraint of Eq.~\eqref{eq:det-constraint}. A detailed proof of this fact may be found in M. Mancini's M.S. Thesis \cite{Mancini:2022}. In summary, imposing the rank-deficiency of the \(2\times2\) matrix \(\bA_i^T\bQ^*\bA_i\) produces the same type of constraint as zeroing a measure of the distance between the points imaged by the camera and the projection of the orbital path on its image plane.

\subsection{From Disk Quadric to Orbital Elements}
Once we have obtained the disk quadric \(\bQ^*\), recovering the classical orbital elements is straightforward. This is essentially the reverse mapping to the relationship described by Eq.~\eqref{eq:qstar}. Given the disk quadric \(\bQ^*\), we can directly compute the ellipse semi-minor axis \(b\) as
\begin{equation}
b=\sqrt{-1 / Q^*_{33}}
\end{equation}
and the focal distance $c$ as
\begin{equation}
c = b^2 \|\bQ_{1:3,4}\|
\end{equation}
Consequently, the semi-major axis and the eccentricity exploiting the relations \(a=b^2+c^2\) and \(e=c/a\). 
Then, we can find \(\bp\) as
\begin{equation}
\bp=\frac{\bQ_{1:3,4}}{\|\bQ_{1:3,4}\|},
\end{equation}
and we can write
\begin{equation}
\bq \bq^T=\bQ^*_{1:3,1:3}-\bp\bp^T.
\end{equation}
Using the fact that \(\bq\) is a unit vector, we finally obtain for some $i = 1, 2, 3$ that
\begin{equation}
\bq=\frac{\left(\bq\bq^T\right)_{1:3,i}}{\|\left(\bq\bq^T\right)_{1:3,i}\|}.
\end{equation}
Here \(i\) may be chosen such that the selected column of $\bq \bq^T$ has largest norm, so as to avoid numerical issues. 

Determining \(i\), \(\Omega\) and \(\omega\) from \(\bp\) and \(\bq\) can be done using standard relations that can be found, for example, in \cite{Vallado}.

Determining the true anomaly corresponding to a bearing measurement requires a few more steps. Let \(\bx_0\) and \(\bu_0\) be the observer's position and the observed direction relative to that bearing. The coordinates of the observed point along the orbit are given by
\begin{equation}\label{eq:position_on_the_orbit}
    \brr_0=\bx_0+\lambda\bu_0
\end{equation}
for some \(\lambda\in\mathbb{R}\). The corresponding coordinates in \(\mathbb{P}^3\) are:
\begin{equation}
    \overline{\brr}_0 \, \propto \, \begin{bmatrix}
    \brr_0\\
    1\\
    \end{bmatrix}
\end{equation}
Since the satellite's position must be a point of the orbital plane \(\boldsymbol{\beta}\), with 
\begin{equation}
\boldsymbol{\beta}\, \propto \, \begin{bmatrix}
w_1\\
w_2\\
w_3\\
0\\
\end{bmatrix}
\end{equation}
the following relationship allows us to determine \(\lambda\)
\begin{equation}
    \overline{\brr}_0^T\boldsymbol{\beta}=0
\end{equation}
which gives
\begin{equation}
\lambda=-\frac{\bx_0^T\bw}{\bu_0^T\bw}
\end{equation}
Substituting this expression for \(\lambda\) inside Eq.~\eqref{eq:position_on_the_orbit}, we can finally solve for the true anomaly corresponding to the bearing as the angle between the periapsis direction, enclosed in \(\bg\) (or \(\bp\)), and the position vector \(\brr_0\):
\begin{equation}
    \nu_0= \operatorname{atan2} \left(\frac{\|\bg\times\brr_0\|}{\|\bg\|\|\brr_0\|},\frac{\bg\cdot\brr_0}{\|\bg\|\|\brr_0\|}\right)
\end{equation}

Note that \(\lambda\) is also the range between the observer's position and the satellite's observed position. If desired, it is possible to correct for the light time-of-flight using \(\lambda\), together with the time of a single measurement.

\section{Orbit determination via polynomial system solving}

\subsection{Formulation as a polynomial system of equations}

If we take a scaling factor of $1$ in the parameterization of $\bQ^*$ given in Eq.~\eqref{eq:qstar2}, the seven unknown orbital parameters $\bw , \bg , b$ will be solutions to a system of $2+n$ equations given by Eqs.~\eqref{eq:constraints-wg} and~\eqref{eq:det-constraint} for $i=1, \ldots , n.$
For \emph{sufficiently generic} observations, we expect that when $7=2+n,$ or $n=5,$ this system of equations will have finitely many solutions.
In fact, we expect the number of solutions to be \emph{constant} if we count over the complex numbers.
Such a ``number conservation" principle may be viewed as a generalization of the fundamental theorem of algebra, which treats the case of a single polynomial in a single unknown.
A formal statement in the language of algebraic geometry may be found, for instance, in Ref.~\cite[Theorem 2.29]{ShafarevichVol1}.

In the preceding paragraph, the precise meaning of the phrase ``sufficiently generic" is that the vector of all observations 
$ \begin{bmatrix} \bx_1^T & \cdots & \bx_5^T \,&\, \bu_1^T & \cdots & \bu_5^T \end{bmatrix} \in \RR^{30}$ lies outside of an appropriately-defined \emph{discriminant locus}.
This discriminant locus is analogous to the commonly-known discriminant of a quadratic equation, but very difficult to describe explicitly (e.g., see \cite[Sec.~3.1]{SturmfelsCBMS} for an example with two unknowns.)

To see an example which is \emph{not} sufficiently generic, we may consider any $5$ observations from any perfectly-circular orbit.
In this case, the true orbit satisfies $\bg = 0$ and the corresponding solution is \emph{singular}; that is, the $7\times 7$ Jacobian matrix of the system evaluated at this solution is rank-deficient.
Such a singular solution may be difficult to estimate accurately with standard numerical methods.
The same is true for the nearly-singular solutions arising from the practical case of nearly-circular orbits.

On the other hand, if we enforce the constraint of a circular orbit by requiring that $\bg = 0$, then $n=3$ observations suffice to recover the orbit up to finitely many possibilities.
Under this circular model, we obtain another system of polynomial equations, from which we may compute the remaining orbital parameters $\bw , b$.
Such a model has the advantage of needing fewer observations, and may potentially give a reasonable approximation of the true orbit in the nearly-circular case.

We summarize the two different orbit models, the associated polynomial systems, and how many solutions they have for generic data in Proposition~\ref{prop:root-count}.

\begin{proposition}\label{prop:root-count}
\begin{itemize}
\item[1.] For $n=5$ generic observations $(\bx_i, \bu_i)_{i=1}^5,$ the elliptical model given by Eqs.~\eqref{eq:constraints-wg} and~\eqref{eq:det-constraint} for $i=1, \ldots , 5$ has a total of $\bf{66}$ complex solutions in the unknown matrix $\bQ^*,$ each lifting to $4$ solutions $(\pm \bw , \bg , \pm b)$ in terms of the parameterization~\eqref{eq:qstar2}.
\item[2.] For $n=3$ generic observations $(\bx_i, \bu_i)_{i=1}^3,$ the circular model given by Eq.~\eqref{eq:det-constraint} for $i=1, \ldots , 3$, plus the additional constraints
\begin{equation}
\label{eq:constraints-wg-circular}
\bw^T \bw = 1, \phantom{ff}
\bg = 0,
\end{equation}
has a total of $\bf{12}$ complex solutions in the unknown matrix $\bQ^*,$ each lifting to $4$ solutions $(\pm \bw , 0 , \pm b)$ in terms of the parameterization~\eqref{eq:qstar2}.
\end{itemize}
\end{proposition}

The claims appearing in Proposition~\ref{prop:root-count} regarding the number of solutions may be readily verified using any one of the standard methods for solving polynomial systems, such as Gr\"{o}bner bases (see~\cite[Ch.~2]{CLO15} for an overview), or polynomial homotopy continuation~\cite{Sommese:2005}.
In our experiments, we solve the systems associated to either model using an implementation of the latter method provided by the software package \texttt{NAG4M2}~\cite{Ley11} in the computer algebra system \texttt{Macaulay2}~\cite{M2}.
We give an overview of homotopy continuation in the next subsection.

\begin{remark}\label{remark:reality}
For different choices of real parameters $(\bx_i, \bu_i)_{i=1}^n,$ the systems appearing in Proposition~\ref{prop:root-count} may have one or more \emph{real} solutions.
For either of our two models, the precise notion of ``real solution" turns out to depend on which quantities are considered as unknowns.
For example, the real disk quadric given by
\[
\bQ^* = \begin{bmatrix}
3/2 & 12 & 0 & 1 \\
1/2 & 3/2 & 0 & 0\\
0 & 0 & 1 & 0\\
1 & 0 & 0 & -1
\end{bmatrix}
\]
can only be lifted to the complex-valued parameters
\[
\bw^T = (\pm i/\sqrt{2}) \, \begin{bmatrix}
1 & 1 & 0 
\end{bmatrix},
\phantom{f}
\bg^T = \begin{bmatrix}
1 & 0 & 0
\end{bmatrix},
\phantom{f} 
b = \pm 1.
\]
\end{remark}

\subsection{Solving polynomial systems with parameter homotopies}

Numerical homotopy continuation is a general method which can be used to solve polynomial systems, such as those appearing in Proposition~\ref{prop:root-count}.
The essential idea underlying the method is as follows: since the roots of polynomials vary continuously with their coefficients, we can use a system whose solution set is known (the \emph{start system}) to solve some other system in the same class of systems (the \emph{target system}) by estimating how the solutions change (\emph{path-tracking}) as we deform one system into another.
We use a short example to illustrate the main ideas.

\begin{example}\label{ex:homotopy-example}
To solve the target system in two variables $\bX = \begin{bmatrix} x & y \end{bmatrix}^T$ given by
\[
\bF_1 (\bX) = \begin{bmatrix}
-x^3 + 2\, x + 1 & - y^2 + x + 1
\end{bmatrix}^T = 0,
\]
we may use the well-known \emph{total-degree} start system~\cite[Sec.~8.4.1]{Sommese:2005}
\[
\bF_0 (\bX) = \begin{bmatrix}
x^3 -1 & y^2 -1
\end{bmatrix}^T = 0,
\]
and the straight-line homotopy
\[
\bH (\bX; t ) = (1-t) \, \bF_0 (\bX) +  t\, \bF_1 (\bX) = 0,
\]
The $6$ start solutions have the form $\begin{bmatrix}e^{2 \pi i k / 3} & \pm 1 \end{bmatrix}^T$ for $k=1,2,3.$
Each determines an initial value for a \emph{solution path} $\bX (t)$, defined for $t$ near $0$, which satisfies $H(\bX (t);t) = 0.$
For this particular homotopy, each solution path is defined for all $t\in [0,1],$ and each value $\bX (1)$ gives one of the six target solutions.
For instance, if $\bX (0) = \begin{bmatrix}1& 1\end{bmatrix}^T$, then $\bX(1) \approx \begin{bmatrix}.246, -.712\end{bmatrix}^T.$
\end{example}

The choice of the start system is an important factor when implementing any homotopy continuation method.
Usually, we want a start system that is general enough to solve any possible target system coming from a specific application.
Another important factor when choosing a start system is its \emph{specificity}, or the number of start solutions.
Different choices of start systems are compared in Ref.~\cite[Ch. 8]{Sommese:2005}, where a basic tradeoff is identified: a start system that is easy to describe and solves a large class of systems will typically require tracking more paths.
For example, if the total-degree start system is used to solve an instance of the elliptical model, this requires tracking $4^5 \times 2^2 = 4096$ paths. This should be compared with the optimal number of $66$ established by Proposition~\ref{prop:root-count}.
If the total degree homotopy was to be used to solve the elliptical model, then we would need to track $4096 - 66 = 4030$ \emph{divergent} solution paths with $\bX (t) \to \infty $ as $t \to 1^-.$ 

In contrast to the total degree homotopy, \emph{parameter homotopies}~\cite{MR977815} allow us to track, under reasonable assumptions, the optimal number of paths.
Since ``most" instances of the systems in Proposition~\ref{prop:root-count} will have the same number of solutions, any randomly-chosen instance may, in principle, be chosen as a start system.
Parameter homotopies may be used in a general setting where we have a system $\bF (\bX ; \bP ) = 0$, with as many equations as unknowns, which depends polynomially both on the unknowns $\bX $ and certain \emph{parameters} $\bP$ depending on the observations.
For the systems of interest to us, there is some flexibility in how these parameters are chosen.
One simple choice is that the target parameters $\bP_1 \in \RR^{8 n}$ consist of all dual coordinates of all $n$ lines---that is, $\bP_1 \in \RR^{40}$ for the elliptical model, and $\bP_0 \in \RR^{24}$ for the circular model.
Alternatively, using Eq.~\eqref{eq:algebraic-Ai}, we may directly encode the observer and bearing vectors into a parameter vector $\bP_1 \in \RR^{6n} $---that is, $\bP_1 \in \RR^{30}$ for the elliptical model, and $\bP_1 \in \RR^{18}$ for the circular model.
The following description of parameter homotopies applies equally well to either choice.

In general, the parameter values $\bP_1$ encoded by a set of observations  specify the target system $\bF (\bX ; \bP_1) = 0$ of the parameter homotopy.
If there are $m$ parameter values, i.e., $\bP_1\in \RR^m$, then the start system should also be a $m\times 1 $ vector $\bP_0.$
For the start system $\bF ( \bX ; \bP_0) =0,$ we are given a pre-computed set of $d\in \{ 12, 66 \}$ complex solutions $\bX_1 (0), \ldots , \bX_d (0)$ whose coordinates are the unknown orbital elements which are pairwise-inequivalent up to the sign-symmetries described in Proposition~\ref{prop:root-count}. 
To extend these start solutions to \emph{solution paths} $\bX_1 (t), \ldots , \bX_{d} (t),$ we ``deform" the start system into the target system via the parameter homotopy
\begin{equation}\label{eq:homotopy}
\bH (\bX ; \bP , t) = \bF (\bX ; t \bP_1 + (1-t) \bP_0 ) = 0.
\end{equation}
Each solution path $\bX (t)$ is an implicit function of $t$ satisfying Eq.~\eqref{eq:homotopy} and the nonlinear ODE system
\begin{equation}\label{eq:solution-path}
\displaystyle\frac{d \bH }{d \bX}  \,
\displaystyle\frac{d \bX}{dt} + \displaystyle\frac{d \bH}{dt} = 0.
\end{equation}

This ODE, together with one of the start solutions $\bX_i (0)$, gives an initial value problem for an unknown solution path $\bX_i (t)$ satisfying Eq.~\eqref{eq:homotopy}.
Numerical integration methods allow us to estimate $\bX_i (t)$ for $t\in [0,1].$
In practice, we approximate a solution path using numerical/predictor corrector methods.
If $\bX_i (t)$ is known within some tolerance for some $t\in [0,1],$ then a ``predictor step" (typically the standard fourth-order Runge-Kutta method) uses Eq.~\eqref{eq:solution-path} to find an initial estimate for $\bX_i (t + \Delta t)$ for some step-size $\Delta t.$
This estimate is subsequently refined by a ``corrector step" using one or more iterations of Newton's method.

A key property of parameter homotopies is that, under mild hypotheses, they are \emph{globally convergent with probability-one.} 
The following key properties of the parameter homotopy in Eq.~\eqref{eq:homotopy} are consequences of a more general parameter continuation theorem~\cite[Theorem 7.1.1]{Sommese:2005}.
We specialize this general result to our case of particular interest.
For sufficiently generic (and hence, \emph{almost all}) $(\bP_0,\bP_1),$ we have:
\begin{enumerate}
\item Each start solution $\bX_i (0)$ extends to a solution path $\bX_i (t)$ which is smooth for all $t \in [0,1].$
\item Every solution to the target system may be obtained up to sign-symmetry from the \emph{endpoint} $\bX_i (1)$ of some solution path $\bX_i (t).$
\end{enumerate}

At this point, it is worth pointing out key differences between homotopy continuation, a \emph{global} root-finding method, from a more standard \emph{local} approach like Newton's method.
In Newton's method, we would pick a single ``start solution" $\bX$, then compute iterates $\bX \gets \bX - \bF(\bX ; \bP_1) $ until some convergence condition is satisfied, yielding a single solution to the target system $\bF (\bX ; \bP_1)=0.$
In practice, this may suffer if the initial guess for $\bX$ is not sufficiently close to a target solution.

In homotopy continuation, ``start solutions" instead refer to solutions to the start system $\bF (\bX ; \bP_0) =0$ which have been pre-computed.
There is no requirement that these start solutions be ``close" to solutions of the target system, nor that the start and target parameters be close in the space of all parameters.
As long as the pre-computed start parameters $\bP_0 \in \CC^{m}$ are \emph{sufficiently generic}, all start solutions can be numerically continued from $t=0$ to $t=1.$
Thus, we can compute \emph{all} solutions to the target system $\bF (\bX ; \bP_1) =0,$ provided that the target parameters $\bP_1$ are also sufficiently generic.

Usually, choosing random complex numbers for the start parameters $\bP_0$ will be sufficient to ensure that all isolated solutions of the target system specified by $\bP_1$ can be computed.
As such, the choice of start parameters, provided that they are sufficiently random, may be expected to have relatively little impact on the accuracy of the computed target solutions.
Here, ``accuracy'' describes the distance between a true target solution and the approximation that is computed by homotopy continuation.
In the absence of measurement noise, the truth is usually recovered to within machine precision. With measurement noise, the accuracy is dependent on the amount of noise (i.e., more noise results in poorer accuracy).  
Regardless of the presence (or magnitude) of noise, there is no clear, simple criterion for picking a ``good" start system that may lead to more accurate results.
However, there are some properties of the start parameters that we can aim to optimize.
For instance, if the Jacobian matrices $\displaystyle\frac{d \bH }{d \bX} (\bX_i (0) ; \bP_0)$ are well-conditioned for each of the start solutions $\bX_i(0)$, then for each solution path we can expect the predictor/corrector steps to be more accurate when $t$ is near $0.$ 
With this in mind, our choice of start system was based on generating several different sets of parameters $\bP_0$ at random, with coordinates drawn uniformly from the complex unit circle, computing a complete set of solutions for each system, and selecting the system whose maximum condition number over all solutions was smallest.
To solve the parametric systems in each model for an initial set of parameters, we used an approach based on monodromy, which works well in tandem with parameter homotopies and can naturally exploit the four-fold symmetry in solutions.
We refer to Refs.~\cite{MR3984062,https://doi.org/10.48550/arxiv.2105.04460,https://doi.org/10.48550/arxiv.1612.08807} for more details about this approach.

In our implementation, we use Eq.~\eqref{eq:numerical-Ai} to determine the parameters representing the dual coordinates of each line-of-sight observation.
Before solving, we rescale all distance measurements into units of earth-radii, which has the typical effect of making the entries of $\bQ^*$ comparable in magnitude.
We use the default path-tracker settings in \texttt{NAG4M2}, except that a minimum stepsize of $\Delta t = \num{1e-16}$ is used.
Additionally, we handle (infrequent) cases of path failure for $t\approx 1$ using Cauchy's endgame for estimating singular solutions~\cite[Ch.~10]{Sommese:2005}.

\subsection{An over-determined formulation}\label{sec:over-determined}

For $n>5$ observations, an exact solution satisfying all $n+2$ constraints of Eqs.~\eqref{eq:det-constraint} and~\eqref{eq:constraints-wg} typically will not exist.
However, we can still aim to minimize some cost function $J$ that depends on all $n$ measurements.
One simple choice based on Eq.~\eqref{eq:det-constraint} is a sum of squared equation residuals,

\begin{equation}\label{eq:cost-function}
J (\bQ^*) = \displaystyle\sum_{1\le i \le n} \left(det(\bA_i\,\bQ^*\bA_i)\right)^2.
\end{equation}

Our task is now to minimize $J (\bQ^*)$ while enforcing the constraints from Eq.~\eqref{eq:constraints-wg}. 
To solve this constrained optimization problem, we use the method of Lagrange multipliers to compute all complex-valued critical points.
For $n>5$ generic observations, the number of critical points turns out to be $4244 = 4 \times \textbf{1061}$ critical points.
A solution of minimum cost among the physically-plausible critical points then provides an estimate for the unknown orbit.

To obtain polynomial equations from Eqs.~\eqref{eq:cost-function} and~\eqref{eq:constraints-wg}, we may set $\beta = -1/b^2$ and write down the (homogenized) Lagrangian function,
\begin{equation}\label{eq:lagrangian2}
\mathcal{L} (\lambda_0, \lambda_1, \lambda_2, \bw , \bg , \beta ) = \lambda_0 J (\bg , \bw, \beta ) + \lambda_1 (\bw^T\bw-1) + \lambda_2 (\bw^T\bg).
\end{equation}
The first-order optimality conditions then read
\begin{equation}\label{eq:lagrangian}
\nabla_{\lambda_1,\lambda_2,w_1,...,g_3,\beta } \mathcal{L} = \textbf{0}_{9\times1}.
\end{equation}
Imposing a generic affine-linear equation
\begin{equation}\label{eq:affine-linear}
c_0 \lambda_0 + c_1 \lambda_1 + c_2 \lambda_2 + c_3 = 0,
\end{equation}
we obtain a system of 10 equations (from Eqs.~\eqref{eq:lagrangian} and~\eqref{eq:affine-linear}) in 10 unknowns, depending on the observations $\bA_1 , \ldots , \bA_n$ and the new parameters $c_0, \ldots , c_3.$
As before, the solutions obey a 4-fold symmetry, and we may track 1061 paths using parameter homotopies to compute all solutions for generic parameter values.
With this general setup, one may make the usual choice of parameters $(c_0, c_1, c_2, c_3) = (1, 0, 0, -1)$ to obtain $\lambda_0=1$.
However, to improve numerical stability~\cite[Sec 4.7]{Bertinibook}, we choose these four parameters in our experiments uniformly at random from the unit $3$-sphere in $\RR^4$.

\section{Experiments}
Throughout our experiments, we consider an orbit with known parameters, the \emph{true} solution, represented as the disk quadric $\bQ^*.$
All distances are computed in units of earth-radii so that the entries of $\bQ^*$ are comparable in magnitude.
To estimate the true solution from the $d$ solutions computed by the homotopy continuation solver, represented by disk quadrics $\widehat{\bQ_1^*}, \ldots , \widehat{\bQ_d^*}$, we take the closest solution

\begin{equation}\label{eq:Q-estimate}
\widehat{\bQ^*} = \displaystyle\argmin_{\widehat{\bQ_1^*}, \ldots , \widehat{\bQ_d^*}} \| \widehat{\bQ_i^*} - \bQ^* \|_2,
\end{equation}
where $\| \bullet \|_2$ denotes the $\ell_2$ Hermitian vector norm on the complex vector space of $4\times 4$ symmetric matrices,
\begin{equation}
\| \bQ^* \| = \sqrt{\displaystyle\sum_{1 \le i \le j \le 4} \lvert Q_{i,j}^*\rvert ^2}.
\end{equation}

We measure the error in the estimate of $\bQ^*$ given in~\eqref{eq:Q-estimate} with an absolute error:

\begin{equation}\label{eq:Q-err}
\Qerr = \| \widehat{\bQ^*} - \bQ^* \|_2,
\end{equation}

Additionally, we will compare the true orbit's classical orbital elements $(a,e,i,\Omega , \omega )$ to their estimates $(\widehat{a}, \widehat{e}, \widehat{i}, \widehat{\Omega}, \widehat{\omega})$.
Note that since $\bQ^*$ remains unchanged with respect to variations of the sign of $\bw$, we will recover the longitude of the node \(\Omega\) and the argument of periapsis \(\omega\) with an ambiguity of \(\pi\), which can be easily resolved if the direction of motion is known.

We use the following error measures:
\begin{equation}
\begin{split}
\aerr &=  \widehat{a} -a \\
\eerr &= \widehat{e}-e \\
\ierr&= \theta (\widehat{i},i) \\
\longerr &=  \theta (\widehat{\Omega},\Omega)\\
\argerr &= \theta (\widehat{\omega},\omega)
\end{split}
\end{equation}
where $\theta $ measures the signed difference between the two angles.

\subsection{Scenario 1: a nearly-circular orbit}\label{sec:scenario1}

Since most orbits of interest are nearly circular, we chose a nearly-circular orbit as a first test case for the performance of this purely geometric method. Specifically, we took inspiration from the orbit traveled by the satellite of the AQUA mission \cite{Parkinson:2003}.
Its orbital elements can be found in Table~\ref{tab:AQUAelements}. 

\begin{table}[!ht]
\caption{Orbital elements of a satellite in nearly-circular orbit.}\label{tab:AQUAelements}
\centering
\begin{tabular}{|c|c|c|c|c|}
\hline
\textit{a} & \textit{e} & \textit{i} & $\Omega$ & $\omega$ \\ \hline
 7080.6 km & 0.0015 & $98.20^{\circ}$ & $95.21^{\circ}$ & $120.48^{\circ}$\\
\hline
\end{tabular}
\end{table}

We obtained simulated observer data considering observations gathered by different ground stations.
Ten observers are given in geocentric coordinates, in units of Earth radii, by 
\[
\begin{bmatrix} \,
\bx_1 & \cdots & \, \bx_{10} \,
\end{bmatrix} \approx
\left[\begin{smallmatrix}
      {\phantom{-}.238}&{\phantom{-}.327}&{\phantom{-}.399}&{\phantom{-}.179}&{\phantom{-}.238}&-{.018}&-{.006}&-{.089}&-{.128}&-{.186}\\
      -{.733}&-{.913}&-{.789}&-{.980}&-{.733}&{\phantom{-}.948}&{\phantom{-}.997}&{\phantom{-}.989}&{\phantom{-}.672}&{\phantom{-}.247}\\
      {\phantom{-}.637}&{\phantom{-}.243}&{\phantom{-}.467}&{\phantom{-}.087}&-{.637}&-{.319}&{\phantom{-}.078}&{\phantom{-}.122}&{\phantom{-}.729}&{\phantom{-}.951}\\
      \end{smallmatrix}\right].
\]

A corresponding set of unit-length bearings is obtained from known points along the orbit:
\[
\begin{bmatrix} \,
\bu_1 & \cdots & \, \bu_{10} \, \end{bmatrix} \approx \left[\begin{smallmatrix}
      -{.226}&-{.587}&-{.443}&-{.381}&-{.343}&{\phantom{-}.281}&-{.288}&-{.014}&{\phantom{-}.619}&{\phantom{-}.722}\\
      -{.755}&-{.495}&-{.451}&-{.486}&-{.645}&-{.958}&{\phantom{-}.053}&{\phantom{-}.907}&{\phantom{-}.776}&-{.624}\\
      -{.616}&-{.641}&-{.775}&-{.786}&{\phantom{-}.683}&{\phantom{-}.060}&-{.956}&-{.421}&-{.122}&{\phantom{-}.298}\\
      \end{smallmatrix}\right].
      \]
The true orbit's disk quadric $\bQ^*$ is
\[
\bQ^* \approx \left[\begin{smallmatrix}
      {\phantom{-}.0284}&-{.0885}&{\phantom{-}.1406}&{\phantom{-}.0002}\\
      -{.0885}&{\phantom{-}.9919}&{\phantom{-}.0128}&-{.0007}\\
      {\phantom{-}.1406}&{\phantom{-}.0128}&{\phantom{-}.9797}&{\phantom{-}.0012}\\
      {\phantom{-}.0002}&-{.0007}&{\phantom{-}.0012}&-{.8114}\\
      \end{smallmatrix}\right].
\]

Since the eccentricity of the orbit is  $e=$\num{1.5e-3}, we might expect that the circular model gives a reasonable approximation of the true orbit.
Thus, to obtain a preliminary assessment of the feasibility of using parameter homotopies for this IOD scenario, we considered both the circular and elliptical models.
In this experiment, $100$ random observer-bearing correspondences were sampled, from a total $120 = \binom{10}{3}$ possibilities for the circular model and $252 = \binom{10}{5}$ for the elliptical model, and the corresponding polynomial systems appearing in Proposition~\ref{prop:root-count} were solved.
Mean and standard deviation of the estimated disk quadric and of the orbital parameters for this experiment are reported in Table~\ref{tab:aqua-noiseless}. For the circular model, we do not report the argument of periapsis $\omega$ since it is undefined.

\begin{table}[!ht]
\footnotesize
\centering
\caption{Mean and standard deviation of the errors for the noiseless analysis of the nearly-circular orbit, for the circular and elliptical models.}
\begin{tabular}{lcccc}
\hline 
& \multicolumn{2}{c}{circular} & \multicolumn{2}{c}{elliptical}\\
 & mean & std & mean & std \\
 \hline
$\Qerr$ & \(\phantom{-}2.15 \times 10^{-3}\) & \(5.08 \times 10^{-5\phantom{8}}\) &\(2.11\times 10^{-12}\) & \(4.68 \times 10^{-17}\) \\
$\aerr [km]$ & \(-2.40\) & \(0.92\) &\(2.82\times 10^{-11}\) & \(2.73 \times 10^{-12}\)\\
$\eerr$ & \(-1.50 \times10^{-3}\) & \(1.09 \times 10^{-18}\) & \(1.49 \times 10^{-14}\)& \(1.36 \times 10^{-15}\) \\
$\ierr [deg]$  & \(\phantom{-}2.80\times10^{-2}\) &\(8.22 \times 10^{-3\phantom{8}}\) & \(2.61 \times 10^{-13}\) & \(2.69 \times 10^{-14}\) \\
$\longerr$ [deg]  & $-1.30\times 10^{-2}$ & $6.48 \times 10^{-3\phantom{8}}$ &\(9.15 \times 10^{-14}\) & \(9.69 \times 10^{-15}\)\\
$\argerr$ [deg] & -- & -- & \(1.72 \times 10^{-11}\) & \(4.08 \times 10^{-13}\) \\
\hline 
\end{tabular}
\label{tab:aqua-noiseless}
\end{table}

The homotopy solvers for both models run on the order of less than a second.
The average runtimes for the elliptical model and the circular model of 0.67 seconds and 0.13 seconds, respectively, differ by a factor of roughly $5$. This is due largely to the fact that there are $66/12 = 5.5$ times as many paths to track in the elliptical model vs the circular model.
Additionally, we used a very conservative value of $\num{1e-16}$ for the minimum value of the predictor/corrector stepsize $\Delta t$, so as to guard against potential failures when tracking solution paths.

Under the elliptical model, Table~\ref{tab:aqua-noiseless} shows that we can recover the true orbit to nearly machine precision in the absence of measurement noise.
Errors under the circular model are uniformly higher due to model mismatch, but still small.
We will show later in this section how the circular model may sometimes be preferable in particular situations.

\begin{figure}[!ht]
\centering
\includegraphics[width=0.35\columnwidth,trim=0in 0in 0in 0in,clip]{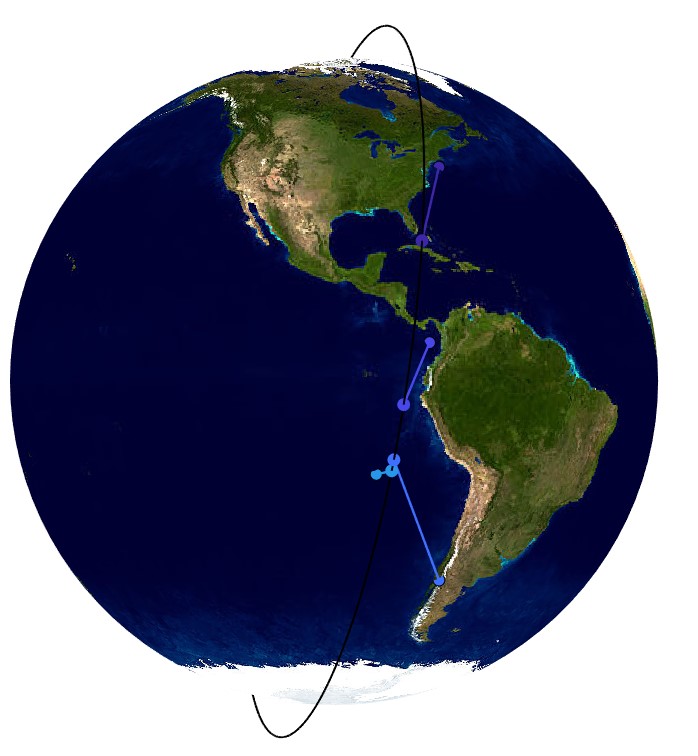}
	\caption{Four of the five observations of the satellite in nearly-circular orbit, including the nearly-coplanar one.}
	\label{fig:configurationAQUA1}
\end{figure}

To illustrate our proposed method of orbit determination on an example from the nearly-circular case, we now focus on a particular set of observers and bearings which appear in Fig.~\ref{fig:configurationAQUA1}, corresponding to columns $1,4,5,6,9$ in the observer and bearing matrices above. 

Note that one of the observations considered is nearly coplanar with the orbital plane, a condition that is singular for the solver, and that is expected to lower its accuracy in presence of noise. 
In the case of ideal observations, under the elliptical model, $44$ of the $66$ complex solutions recovered by the solver give a real-valued disk quadric. However, only $3$ of these real disk quadrics can be lifted to real-valued orbital parameters $(\bw , \bg , b)$ (see Rem.~\ref{remark:reality}.) 
These real solutions are given, modulo the sign ambiguity, by
\[
\begin{split}
\bX_1 &\approx \left[\begin{smallmatrix}
      -{1.00}&-{.033}&-{.022}&{\phantom{-}
.001}&{\phantom{-}
.310}&-{.527}&-{1.136}\\
      \end{smallmatrix}\right],\\
\bX_2 &\approx
\left[\begin{smallmatrix}
      -{.970}&-{.111}&{\phantom{-}
.216}&{\phantom{-}
.033}&-{.068}&{\phantom{-}
.112}&{\phantom{-}
1.171}\\
      \end{smallmatrix}\right],\\
\bX_3 &\approx \left[\begin{smallmatrix}
      -{.986}&-{.090}&{\phantom{-}
.143}&{\phantom{-}
.000}&-{.001}&{\phantom{-}
.001}&-{1.110}\\
      \end{smallmatrix}\right].
\end{split}
\]
The solution $\bX_3$ corresponds to the true orbit.
To rule out the additional real solutions, a sixth line coming from the observation $(\bx_2, \bu_2)$ may be used.
Evaluating the left-hand side of Eq.~\ref{eq:det-constraint} for this line at both solutions $\bX_1$ and $\bX_2$ gives residual errors of $\num{3.9e-3},\num{4.6e-4},$ and $\num{6.2e-16}$, respectively. 
Thus, the solution $\bX_3$ gives the best fit to the sixth observation.

Applying the circular model to this same example (with observations 1, 5, and 6), there are now $6$ real solutions up to sign ambiguity in the nonzero parameters $(\bw , b).$
However, all $5$ original observations may now be used to remove the $5$ extraneous solutions.
Evaluating the residuals of constraints~\ref{eq:det-constraint} on these additional solutions as with the elliptical model allows us to distinguish the correct solution.
Another comparison between the true orbit and the extraneous solutions may be obtained by computing the distance from each of the five estimated orbit points to the origin.
For the three observations used in the homotopy solver these distances all equal $b,$ but these distances may be significantly different for the two unused observations.
For each of the 6 real solutions, we may compute the variance of the set of three distances comprised of $b$ and the unused observations.
One of these values is $\num{1e-9}$, and the rest are on the order of $\num{1e-3}$ or higher. 
This strongly signals the best approximation of the true solution, which is given by
\[
\bX \approx \begin{bmatrix}
-.986 & -.090 & .142 & 0 & 0 & 0 & -1.109
\end{bmatrix}.
\]

\subsection{Scenario 2: an elliptical orbit}\label{sec:scenario2}

As a second test case, we consider an orbit close to that traveled by the Magnetospheric Multiscale Spacecrafts (MMS)~\cite{Fuselier:2014}. 
This is a highly-elliptical orbit, whose orbital elements we report in Table~\ref{tab:MMSelements}.

\begin{table}[!ht]
\caption{Orbital elements of the highly-elliptical orbit.}
\label{tab:MMSelements}
\centering
\begin{tabular}{|c|c|c|c|c|}
\hline
\textit{a} & \textit{e} & \textit{i} & $\Omega$ & $\omega$ \\ \hline
83519.02 km & 0.9082 & 28.50° & 357.84° & 298.22°\\
\hline
\end{tabular}
\end{table}
\begin{figure}[ht!]
\centering
\includegraphics[width=0.5\columnwidth,trim=0in 0in 0in 0in,clip]{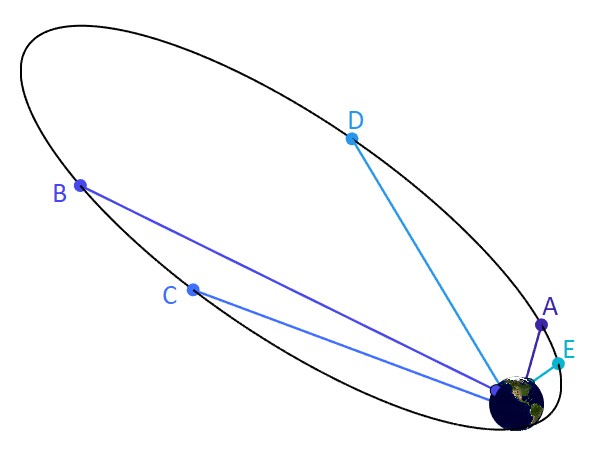}
	\caption{Five observations of the satellite made by three ground stations.}
	\label{fig:configurationMMS}
\end{figure}

In this scenario, 10 observations were obtained with three ground stations on the Earth's surface, and we evaluated the runtime and accuracy on 100 sets of five noiseless observations.
The summary statistics shown in Table~\ref{tab:mms-noiseless} demonstrate that the solver for the elliptical model is efficient and highly accurate in all cases. Its average runtime was of 0.56 seconds.
Note that we do not consider the circular model.

\begin{table}[!ht]
\centering
\footnotesize
\caption{Mean and standard deviation of the solver on noiseless data related to the highly-elliptical orbit.} 
\begin{tabular}{lcc}
\hline
 & \multicolumn{2}{c}{elliptical}  \\
& mean & std \\
\hline
$\Qerr$ & $3.03 \times 10^{-14}$ & $1.02 \times 10^{-15}$ \\
$\aerr [km]$ & $4.72 \times 10^{-9}$  & $1.76 \times 10^{-10}$\\
$\eerr$ & $9.77 \times 10^{-15}$ & $3.14 \times 10^{-16}$\\
$\ierr [deg]$ & $1.22 \times 10^{-13}$ & $3.84 \times 10^{-15}$\\
$\longerr [deg]$ & $-3.05\times 10^{-13}$ & $<10^{-16}$ \\
$\argerr [deg]$ & $-2.03 \times 10^{-13}$ & $8.00 \times 10^{-15}$  \\
\hline 
\end{tabular}
\label{tab:mms-noiseless}
\end{table}

Figure~\ref{fig:configurationMMS} shows five of the 10 line-of-sight observations for the satellite.
We will use these observations in our Monte Carlo study of noisy observations.
Solving the elliptical model yields $11$ real solutions in this case, $10$ of which can be easily excluded by examining the residuals of a sixth line.
With a view towards the model selection problem when the shape of the orbit is not known \emph{a priori}, we also considered what would happen if we solved the circular model for three of these observations.
In this short experiment, four of the resulting 12 solutions turned out to be real.
However, we obtained a strong signal of model mismatch by computing the variances of the estimates of $b$, which resulted in a value greater than 1 for each solution.

\subsection{Noisy observations and comparison with other methods}
To test the robustness of the homotopy solvers to noise under various scenarios, we consider the following noise model.
For each bearing vector $\bu $, a noisy perturbation $\tilde{\bu}$ is given by
\begin{equation}
    \tilde{\bu} = \bu + \epsilon
\end{equation}
where $\epsilon \sim N(0,\bR)$. That is, $\epsilon$ is zero-mean Gaussian noise with covariance following the so-called QUEST measurement model (QMM) \cite{Shuster:1981,Shuster:1989}
\begin{equation}
    \bR = E[ \epsilon \epsilon^T ] = \sigma^2 \left( \bI_{3 \times 3} - \bu \bu^T \right),
\end{equation}
where $\sigma$ is the standard deviation of the bearing error in radians. Note that $\bR$ is a $3 \times 3$ matrix of rank 2, with a null space in the direction of $\bu$. This means that $\epsilon$ lies in the plane normal to $\bu$. If $\epsilon$ is small, then $\tilde{\bu}$ remains a unit vector to first order.
The measured direction will then lie on a cone with axis the true direction, and opening sampled from a normal distribution with standard deviation \(\sigma\).

For all the scenarios analyzed, we ran a Monte Carlo simulation under the noise model described above with 10,000 runs and $\sigma = 1$ arcmin bearing noise, unless differently stated. The objective of the following study is to understand the performance of the solver under different scenarios of observation. Also, we study the role of the elliptical model and the circular model to understand how they can be used to increase the performance in some conditions of observations. 
We compared the results obtained by homotopy continuation with those provided by either the Double-R method \cite{Escobal:1976} or Gauss' method, implemented as described in \cite{Vallado}. To initialize the Double-R iteration, we used an initial guess of three fourth of the true values for the radii of the position vectors. When this value was smaller than the Earth's radius, the sum of the Earth's radius and 1/4 of the true slant range was used. Additionally, in the implementation of these standard IOD solutions, we considered a perfect measurement of the time of the observations. Note that whenever a comparison is made, the same noisy inputs have been given to each solver involved.

\subsubsection{The elliptical orbit}
\begin{figure}[b!]
\centering
\includegraphics[width=1.0\columnwidth,trim=0in 0in 0in 0in,clip]{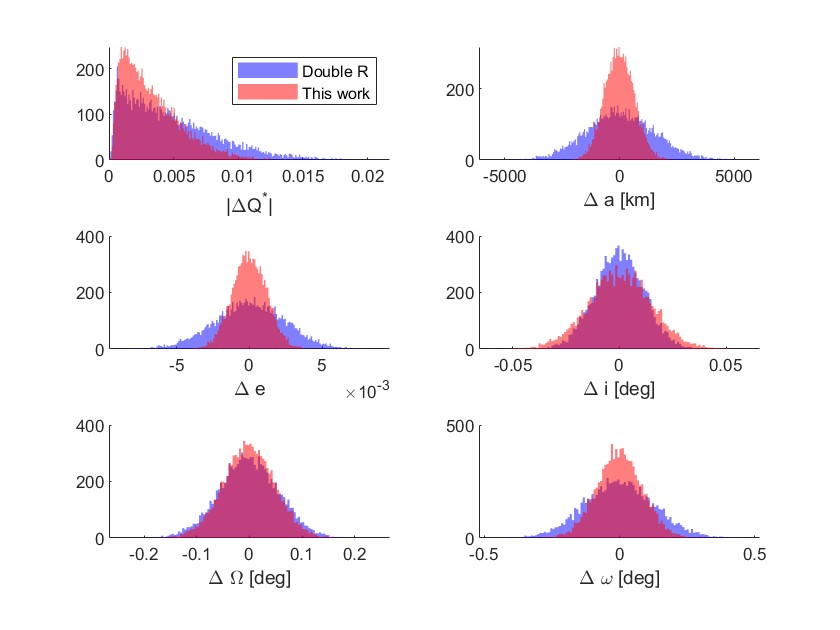}
	\caption{Comparison between the errors produced by the method described in this work and the Double-R iteration method for the set of observations ABE of Fig.~\ref{fig:configurationMMS}. Note that the statistics relative to the other set of observations are presented in Table~\ref{tab:PerformanceMMS}. Results obtained under \(1\) arcmin of noise.}
	\label{fig:compareMMS}
\end{figure}

Consider the observations of the orbit already presented in Fig.~\ref{fig:configurationMMS}. Using these five observations, we can build ten combinations of three observations that are sufficient to find a solution using the Double-R method (together with the times of the measurements). In Table~\ref{tab:PerformanceMMS}, we compare the performance of our solver with the performance provided by Double-R in these 10 cases. From this table, we can see how our method is more accurate in most of the cases, with Double-R failing twice in converging to the true solution. A more thorough comparison between the two methods is given in Fig.~\ref{fig:compareMMS} for the observations ABE. 
\begin{table}[!ht]
\caption{Performance in the estimate of the highly-elliptical orbit, for the observations of Fig.~\ref{fig:configurationMMS}. See Fig.~\ref{fig:compareMMS} for a zoom on the performance for the observations ABE. Results obtained under \(1\) arcmin of noise. Rows with ``F'' indicate observation combinations where Double-R failed to converge.} \label{tab:PerformanceMMS}
\centering
\begin{tabular}{c c c c c c c}
\hline
 observations & IOD method & \begin{tabular}{@{}c@{}}\(\sigma_a\) \\ \([km]\)\end{tabular} &\begin{tabular}{@{}c@{}}\(\sigma_e\) \\ \( \)\end{tabular}&  \begin{tabular}{@{}c@{}}\(\sigma_i\) \\ \([deg]\)\end{tabular} &\begin{tabular}{@{}c@{}}\(\sigma_{\Omega}\) \\ \([deg]\)\end{tabular} & \begin{tabular}{@{}c@{}}\(\sigma_{\omega}\)\\\([deg]\)\end{tabular} \\
 \hline
ABC & Double-R & \(4462\) & \(0.0032\) & \(0.147\) & \(0.230\) & \( 1.256\) \\
ABD & Double-R & \(2227\) & \(0.0056\) & \( 0.040\) & \(0.185\) & \( 0.154\) \\
ABE & Double-R & \(1500\) & \(0.0024\) & \(0.012\)  & \(0.057\)  & \( 0.128\) \\
ACD & Double-R & \(1243\) & \(0.0022\) & \(0.024\)  & \(0.035\)  & \(0.090\)\\
ACE & Double-R & F & F & F & F & F\\
ADE& Double-R & \(13223\) & \(0.0080\) & \(0.073\) & \(0.489\) & \(1.701\) \\
BCD & Double-R & F & F & F & F & F\\
BCE & Double-R & \(2552\) & \(0.0013\) & \(0.159\) & \(0.283\) & \(0.843\) \\
BDE & Double-R & \(685\) & \(0.0017\) & \(0.027\) & \(0.109\)  & \( 0.082\)\\
CDE & Double-R & \(591\) & \(0.0011\) & \(0.022\) & \(0.045\)  &\( 0.065\) \\
ABCDE & This Work & \(646\) & \(0.0012\) & \(0.015\) & \(0.049\)& \( 0.084\) \\
\hline
\end{tabular}
\end{table}

This short analysis shows that the method is competitive with one of the state of the art algorithms for IOD in the case of well-spaced observations, for highly elliptical orbits.

\subsubsection{The nearly-circular orbit}\label{sec:noisyAQUAanalysis}
Given the low altitude of the orbit of the satellite whose orbit was described in Section \ref{sec:scenario1}, we have done a preliminary test on the performance of the method using five different ground stations, one for each observed position. The analysis of this nearly-circular orbit had two objectives. First, understanding whether a nearly-circular orbit is a degenerate solution for the homotopy-based algorithm or not. Also, we wanted to understand the role that the circular model may have in the IOD process, and how it may be used to improve the accuracy of the solution. 

\begin{figure}[ht!]
\centering
\includegraphics[width=0.35\columnwidth,trim=0in 0in 0in 0in,clip]{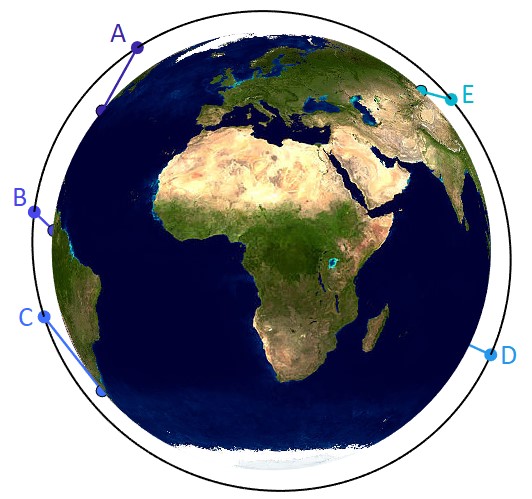}
	\caption{Five observations gathered by five ground stations for the nearly-circular orbit.}
	\label{fig:configurationAQUA5}
\end{figure}

We accomplished the first objective analyzing the performance of the solver for the geometry of observations represented in Fig.~\ref{fig:configurationAQUA5}. Note that these observations are far from being singular. 
      
\begin{figure}[ht!]
\centering
\includegraphics[width=1.0\columnwidth,trim=0in 0in 0in 0in,clip]{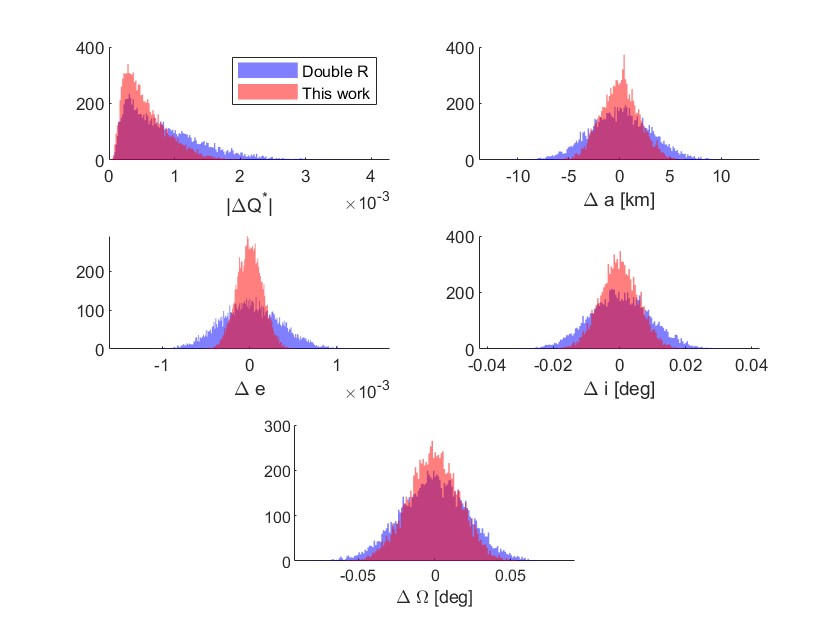}
	\caption{Comparison between the errors produced by the method described in this work and the Double-R iteration method applied to the observations ACE of Fig.~\ref{fig:configurationAQUA5} under $1$ arcmin of noise. A more general comparison can be found in Table.~\ref{tab:PerformanceAQUA}.}
	\label{fig:compareAQUA0}
\end{figure}

With the \textit{elliptical model}, the chosen set of measurements produces the results whose performance is given in Table~\ref{tab:PerformanceAQUA}. Here, we also compare the results with those produced by Double-R, for all the ten possible sets of three observations sampled from the given five. We can see that our method is superior in most of the cases. 
A zoom on the performance for one of the observations is given in Fig.~\ref{fig:compareAQUA0}. 

We can conclude that the method presented in this work proved to be competitive with one of the state of the art algorithms for IOD also for well-spaced observations of a nearly-circular orbit.

\begin{table}[!ht]
\caption{Comparison between the standard deviations of the errors produced by Double-R and by our method for the observations represented in Fig.~\ref{fig:configurationAQUA5}. Figure.~\ref{fig:compareAQUA0} shows in more details the performance relative to the observations ACE. Results obtained under \(1\) arcmin of noise.} \label{tab:PerformanceAQUA}
\centering
\begin{tabular}{c c c c c c}
\hline
 observations & IOD method & \begin{tabular}{@{}c@{}}\(\sigma_a\) \\ \([km]\)\end{tabular} &\begin{tabular}{@{}c@{}}\(\sigma_e\) \\ \( \)\end{tabular}&  \begin{tabular}{@{}c@{}}\(\sigma_i\) \\ \([deg]\)\end{tabular} &\begin{tabular}{@{}c@{}}\(\sigma_{\Omega}\) \\ \([deg]\)\end{tabular} \\
 \hline
ABC & Double-R & \(9.20\) & \(0.00098\) & \(0.034\)  & \(0.032\)  \\
ABD & Double-R & \(0.41\) & \(0.00004\) & \(0.008\)  & \(0.004\)  \\
ABE & Double-R & \(14.25\) & \(0.00132\) & \(0.037\)  & \(0.063\)  \\
ACD & Double-R & \(0.67\) & \(0.00006\) & \(0.009\)  & \(0.012\)  \\
ACE & Double-R & \(3.03\) & \(0.00034\) & \(0.009\)  & \(0.022\) \\
ADE & Double-R &  \(5.78\) & \(0.00018\) & \(0.007\)  & \(0.050\) \\
BCD & Double-R & \(18.85\) & \(0.00221\) & \(0.025\)  & \(0.133\) \\
BCE & Double-R & \(1.99\) & \(0.00030\) & \(0.020\)  & \(0.008\) \\
BDE & Double-R & \(0.15\) & \(0.00006\) & \(0.010\)  & \(0.003\)  \\
CDE & Double-R & \(4.39\) & \(0.00034\) & \(0.055\)  & \(0.073\)  \\
ABCDE & This Work & \(1.86\) & \(0.00015\) & \(0.006\) & \(0.017\) \\
\hline
\end{tabular}
\end{table}

The most interesting analysis, however, was related to understanding the behavior of the solver in the same case studied in the noiseless experiments. We recall that we are using the observations 1, 4, 5, 6, 9 given in section \ref{sec:scenario1}, and that the geometry of the problem is not optimal since one of the LOS is almost coplanar with the orbital plane, as shown in Fig.~\ref{fig:configurationAQUA1}. 

Since the case of coplanar observations is singular for the solver (and for the other classical angles-only IOD solutions \cite{Baker}), we can expect that the accuracy of the solution will decrease. As predicted, the errors in the estimate of the disk quadric, which give an overall idea of the accuracy of the estimate, grew of approximately one order of magnitude with respect the configuration of observations previously analyzed.
For comparison, Double-R was tested over the same noisy inputs: it diverged in all but one of the combinations of observations containing the nearly-singular one.

At this point, since the accuracy obtained with the elliptical model decreased, it makes sense to analyze whether the assumption of a perfectly circular orbit may bring advantages. In the noiseless analysis, we saw that the use of the circular model decreased the accuracy. However, we will show that, in the presence of noise, this model can sometimes improve the solution. 

\begin{figure}[ht!]
\centering
\includegraphics[width=1.0\columnwidth,trim=0in 0in 0in 0in,clip]{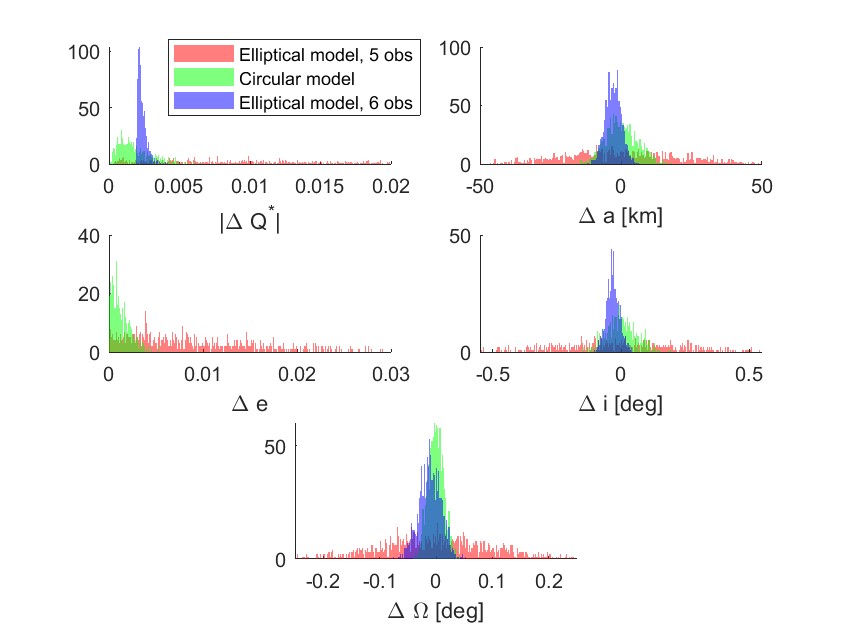}
	\caption{Results obtained with the three models presented in the case of a nearly-singular observation of the nearly-circular orbit. Note that the results relative to Double-R are not reported since Double-R diverged for the three observations used in the circular model. Also, note that the error in eccentricity produced by the circular model is not shown, since in this case is assumed that \(e\equiv0\).}
	\label{fig:AQUAcombined}
\end{figure}

Fixing one of the observations to be the nearly-singular one, and randomly sampling the other two from the remaining, we implemented our circular model. Figure~\ref{fig:AQUAcombined} provides the results of a Monte Carlo simulation with 1000 runs. The circular model is clearly an improvement: the errors are of smaller scale. As a drawback, a bias is introduced, likely due to dynamical model mismatch. 

In order to avoid the bias, then, we came back to the elliptical model, and considered a sixth observation. Using this single additional observation, we can formulate the problem as an optimization problem, as discussed in~\Cref{sec:over-determined}. Now, discerning the true solution among the others is no longer an issue: it will be the one characterized by the smallest value of the cost function. The results showed that the solution obtained using this model reached the accuracy produced by the circular model, with the further advantage of not introducing any bias. This can be appreciated in Fig.~\ref{fig:AQUAcombined}.
The convenience brought by the optimization formulation is evident. A single model (the elliptical one) may be used to solve the problem for any shape of the orbit, overcoming the issue of nearly-singular observations.

\subsubsection{Close observations}\label{sec:shortARCanalysis}
\begin{figure}[ht!]
\centering
\includegraphics[width=.4\columnwidth,trim=0in 0in 0in 0in,clip]{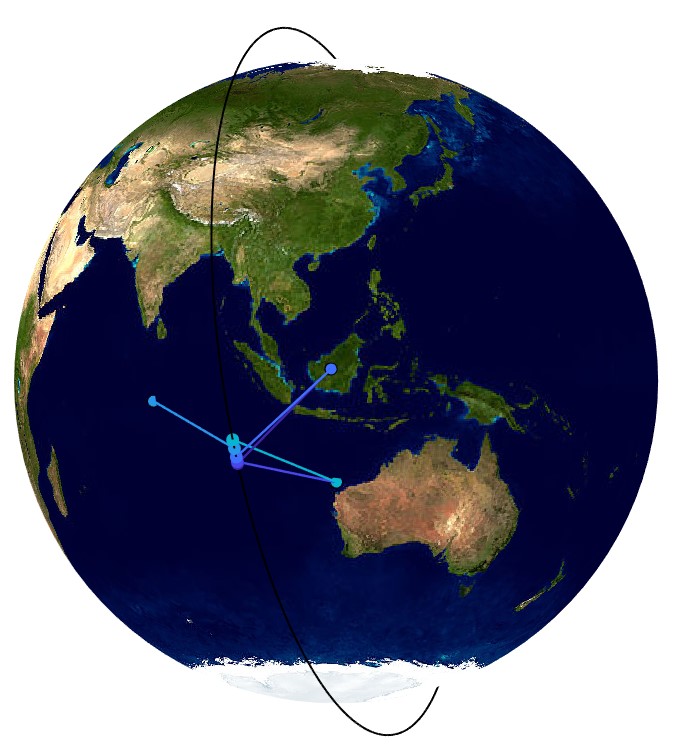}
	\caption{Close observations of the nearly-circular orbit. The observations A-E are ordered from South to North.}
	\label{fig:configurationAQUA2}
\end{figure}

When the observations become closer, the solver shows a decreased tolerance to noise. The performance of our method was tested for observations of the satellite in nearly-circular orbit gathered in an interval of time of about 65 seconds, using three ground stations. The geometry of the observations is represented in Fig. \ref{fig:configurationAQUA2} and the measurements used are the following:
\begin{align*}
\begin{bmatrix} \,
\bx_1 & \cdots & \, \bx_{5} \,
\end{bmatrix} &\approx
\left[\begin{smallmatrix}
      -{.390} & -{.378} & -{.392} & \phantom{-}{.186} & -{.382}\\
      \phantom{-}{.921}&\phantom{-}{.860}&\phantom{-}{.920}&\phantom{-}{.979}&\phantom{-}{.858}\\
      \phantom{-}{.007}&-{.342}&\phantom{-}{.007}&-{.087}&-{.342}\\
      \end{smallmatrix}\right],
      \\
\begin{bmatrix} \,
\bu_1 & \cdots & \, \bu_{5} \, \end{bmatrix} &\approx \left[\begin{smallmatrix}
      \phantom{-}{.609} & \phantom{-}{.746} & \phantom{-}{.633} & -{.868} & \phantom{-}.{699}\\
      \phantom{-}{.348} & \phantom{-}{.639} & \phantom{-}{.372} & \phantom{-}{.267} & \phantom{-}{.621}\\
      -{.712} & \phantom{-}{.188} & -{.679} & -{.419} & \phantom{-}{.356}\\
      \end{smallmatrix}\right].
      \end{align*}

\begin{table}[!ht]
\caption{Standard deviations produced by Gauss' method and our method for the case of close observations shown in Fig.~\ref{fig:configurationAQUA2}. Results obtained under \(1\) arcsec of noise.} \label{tab:PerformanceCloseObs}
\centering
\begin{tabular}{c c c c c c}
\hline
 observations & IOD method & \begin{tabular}{@{}c@{}}\(\sigma_a\) \\ \([km]\)\end{tabular} &\begin{tabular}{@{}c@{}}\(\sigma_e\) \\ \( \)\end{tabular}&  \begin{tabular}{@{}c@{}}\(\sigma_i\) \\ \([deg]\)\end{tabular} &\begin{tabular}{@{}c@{}}\(\sigma_{\Omega}\) \\ \([deg]\)\end{tabular} \\
 \hline
ABC & Gauss & \(97683\) & \(0.18556\) & \(11.695\)  & \(3.500\)  \\
ABD & Gauss & \(3.22\) & \(0.00017\) & \(0.022\)  & \(0.006\)  \\
ABE & Gauss & \(302.59\) & \(0.02664\) & \(1.328\)  & \(0.364\)  \\
ACD & Gauss & \(18.22\) & \(0.00142\) & \(0.093\)  & \(0.021\)  \\
ACE & Gauss & \(545.66\) & \(0.04581\) & \(2.082\)  & \(0.395\)  \\
ADE & Gauss & \(0.83\) & \(0.00008\) & \(0.005\)  & \(0.001\)  \\
BCD & Gauss & \(1.84\) & \(0.00017\) & \(0.012\)  & \(0.003\)  \\
BCE & Gauss & \(134.69\) & \(0.01228\) & \(0.690\)  & \(0.173\)  \\
BDE & Gauss & \(5.03\) & \(0.00068\) & \(0.031\)  & \(0.007\)  \\
CDE & Gauss & \(1.10\) & \(0.00014\) & \(0.007\)  & \(0.002\)  \\
ABCDE & This Work & \(39.94\) & \(0.00324\) & \(0.174\) & \(0.039\) \\
\hline
\end{tabular}
\end{table}

\begin{figure}[ht!]
\centering
\includegraphics[width=1.0\columnwidth,trim=0in 0in 0in 0in,clip]{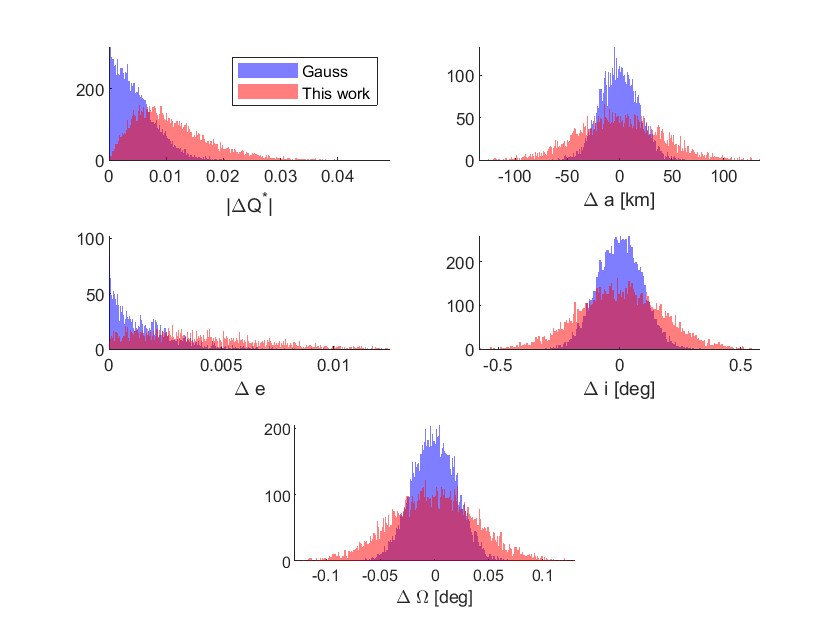}
	\caption{Errors in the  estimate of the nearly-circular orbit under close observations, for the observations ACD of Fig.~\ref{fig:configurationAQUA2} . The time span between the first and the last observation is of 65 seconds. Results obtained under \(1\) arcsec of noise.}
	\label{fig:compareShortArc}
\end{figure}

In this case, for \(\sigma=1\) arcmin the results provided by the solver are poor. Under \(\sigma=1\) arcsec, instead, they are reasonable. Note that Double-R was not able to converge in this case. On the other hand, Gauss's method for angles-only IOD \cite{Gauss:1809} works in all the ten possible configurations of three observations extracted from the given five. In Table ~\ref{tab:PerformanceCloseObs} we compare the performance of the two methods, while Fig.~\ref{fig:compareShortArc} provides the histograms relative to the observations ACD, where Gauss' solver shows a slightly better accuracy than ours. On the overall, we can comment that Gauss' method and our method have comparable performance, without anyone prevailing over the other.

\subsection{The hyperbolic case}
\begin{figure}[ht!]
\centering
\includegraphics[width=0.6\columnwidth,trim=0in 0in 0in 0in,clip]{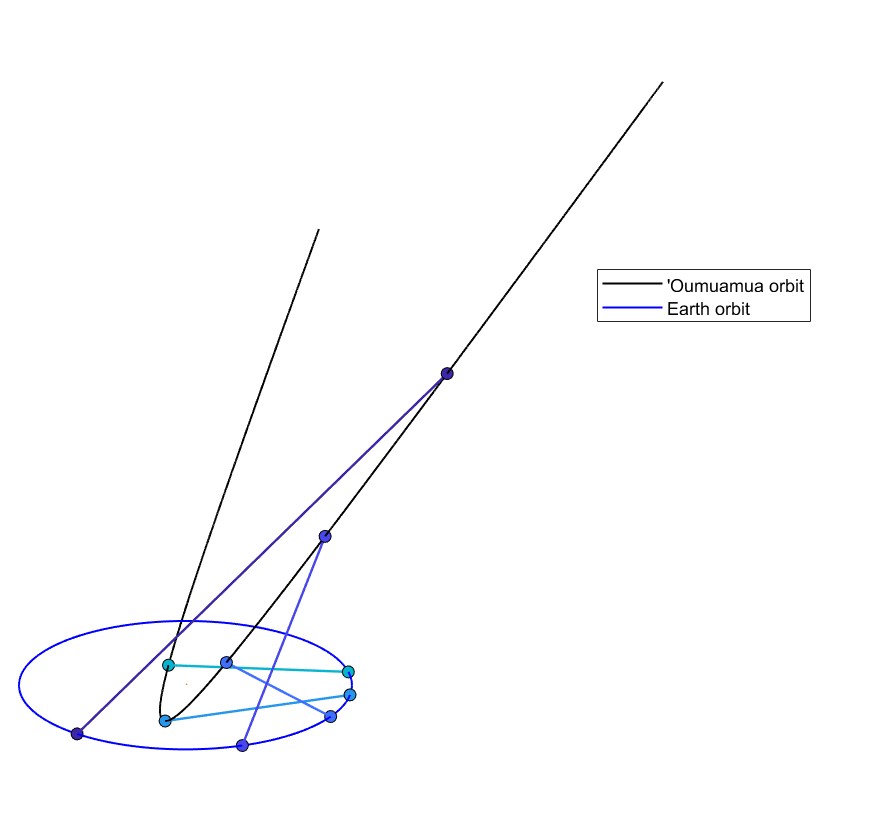}
	\caption{Simulated observations for the `Oumuamua orbit.}
	\label{fig:configurationOum}
\end{figure}

\begin{figure}[ht!]
\centering
\includegraphics[width=1.0\columnwidth,trim=0in 0in 0in 0in,clip]{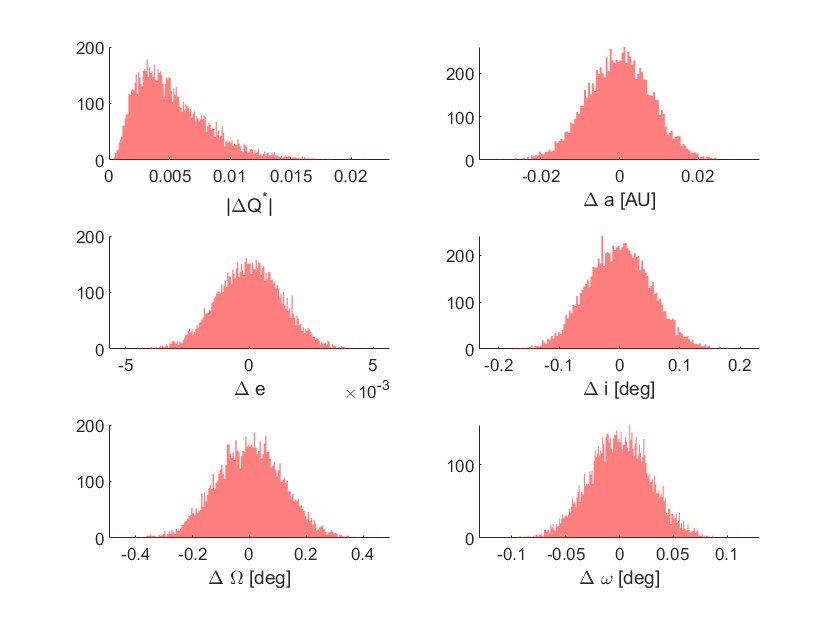}
	\caption{Results obtained for the test case of `Oumuamua, under 1 arcmin of noise, for 10,000 runs.}
	\label{fig:compareOum}
\end{figure}

If the orbit that we aim to recover is an hyperbola, the procedure described above remains valid. Once that the matrix \(\bQ^*\) has been recovered, however, its parameterization in terms of \(\bw, \bg\) and \(1/b^2\) differs from the previous for a sign:
\begin{equation}
    \bQ^* \, \propto \, \begin{bmatrix}
    -\textbf{I}_{3\times3}+\bw\bw^T & \bg\\\bg^T & (-1/b^2)
    \end{bmatrix}
\end{equation}
The hyperbolic formulation has been analyzed recovering the orbit of the first interstellar object ever sighted, `Oumuamua \cite{oumuamua2019natural}. This surprisingly elongated object, whose planetary system of origin is unknown, was discovered in 2017 and has left the Solar system traveling on its hyperbolic orbit. Its heliocentric orbital elements are shown in Table~\ref{tab:OumOrbitalElements} and the five observations that have been simulated are shown in Fig.~\ref{fig:configurationOum}. The performance of the method is represented in Fig.~\ref{fig:compareOum}. Note that the Double-R method cannot be used for comparison since in some of the cases the distance between the observer and the Sun is greater than the distance between the asteroid and the Sun. 

\begin{table}[!ht]
\caption{Orbital elements of the `Oumuamua asteroid.}\label{tab:OumOrbitalElements}
\centering
\begin{tabular}{|c|c|c|c|c|}
\hline
\textit{a} & \textit{e} & \textit{i} & $\Omega$ & $\omega$ \\ \hline
 \(-1.9034 \times 10^8\) km & 1.20 & $122.74^{\circ}$ & $24.60^{\circ}$ & $241.81^{\circ}$\\
\hline
\end{tabular}
\end{table}

\newpage

\section{Conclusion}

This work develops a purely geometric method for solving the angles-only initial orbit determination (IOD) problem under the assumption of Keplerian motion.
Our method is unique in comparison to existing angles-only methods which all, to the best of our knowledge, require time and some form of orbit propagation. 

Experiments demonstrate that our method achieves accuracy that is comparable to, and sometimes better than, one of the standard IOD methods (Double R).
An attractive feature of using homotopy continuation to solve for the unknown conic $\bQ^*$ in Eq.~\eqref{eq:det-constraint} is that it gives a truly global method---no initial guess is required whatsoever.
In an additional set of (unreported) experiments, we witnessed for some configurations that the Double-R iteration may have a strong dependence on the initial guesses for the radii.
Moreover, poor initialization in general may prevent this method from converging at all.
Thus, our approach may be used not only as a standalone IOD solver, but also shows potential as a method for providing initial guesses for parameters used by other methods.
Such an initialization scheme may increase the accuracy or probability of convergence for these other existing methods. 

It is also important to address the model selection problem of deciding between circular and elliptical model when the orbit shape is unknown \emph{a priori}. The analyses made so far allowed us to identify the following strategy. The elliptical model can in general be used, independently of the shape of the closed orbit, and it behaved reliably in providing a solution. If its solution is an orbit with low eccentricity, we can consider using the circular model to provide another estimate of the orbit. In general, the solution obtained with the circular model is not necessarily more accurate than that given by the elliptical model. However, when some of the observations approach the singular configuration of coplanarity with the orbit, we experienced a decrease in the performance of the elliptical model, which made the circular model behave better. Nonetheless, when more than five observations are available, setting the IOD problem as an optimization problem should be preferred. In fact, this allows us to use a unique model for the solution of the IOD problem, increasing the accuracy and avoiding the introduction of the bias produced by the circular model.

On a similar note, we recall $n=5$ (or $n=3$) observations generally only suffice to determine $\bQ^*$ up to finitely many possibilities, since the systems in Proposition~\ref{prop:root-count} may have several real solutions (with the exact number depending on scenario specifics).
We addressed two simple strategies for ruling out ``false solutions" using additional observations. Moreover, considerations of general type may be used to exclude some of the wrong solutions. For instance, orbits with periapsis lower than the orbited planet's surface, or orbits corresponding to observations that happen in the wrong direction, or that must cross a physical object to occur, can in general be discarded. 

In the case of a hyperbolic orbit, a sign difference in the parameterization of the disk quadric makes the recovered solution non-liftable to real values of the parameter \(b\), when using the parameterization for elliptical orbits. This is a clue that the obtained disk quadric should be analyzed in terms of the hyperbolic parameterization.

Finally, once that the disk quadric has been determined, we can recover all the orbital elements without any ambiguity when the direction of motion is known.

\clearpage

\section{Acknowledgments}
The work of M. Mancini and J. Christian was partially supported by the Air Force Office of Scientific Research under the Space University Research Initiative (grant FA95502210092). The work of T. Duff was supported by the National Science Foundation (NSF) Mathematical Sciences Postdoctoral Research Fellowship (DMS-2103310). The work of A. Leykin was partially supported by NSF DMS-2001267. The authors greatly appreciate the support of our sponsors.

\section{Conflict of interest}
On behalf of all authors, the corresponding author states that there is no conflict of interest.
\bibliography{sn-bibliography}


\end{document}